\documentclass[pra,10pt, twocolumn, aps, superscriptaddress, showpacs, groupaddress, floatfix]{revtex4-1}
\usepackage[T1]{fontenc}
\usepackage{bigints}
\usepackage{amssymb}
\usepackage{graphicx}
\usepackage{amsmath}
\usepackage{dcolumn}
\usepackage{multirow}
\usepackage{hyperref}
\usepackage[normalem]{ulem}
\usepackage{physics}
\usepackage{bm}
\usepackage{comment}
\usepackage{xcolor}
\usepackage[caption=false]{subfig}
\usepackage{siunitx}
\usepackage{dsfont}
\usepackage{bbold}
\usepackage[toc]{appendix}
\usepackage{relsize}
\usepackage{leftidx}
\usepackage{bbm}
\usepackage{stackengine}

\definecolor{lapislazuli}{rgb}{0.15, 0.38, 0.61}
\definecolor{YKblue}{rgb}{0.0, 0.18, 0.65}
\definecolor{carmine}{rgb}{0.81, 0.09, 0.13}
\definecolor{lavender}{rgb}{0.84, 0.79, 0.87}

\hypersetup{
	colorlinks=true,
	linkcolor=YKblue,
	linktoc=page,
	citecolor=blue,
	urlcolor=YKblue
}

\newcommand{\bvr}{\mathbf{r}}
\newcommand{\vk}{\mathbf{k}}
\newcommand{\vv}{\mathbf{v}}
\newcommand{\kF}{k_{\mathrm{F}}}
\newcommand{\vF}{v_{\mathrm{F}}}
\newcommand{\kst}{k^{*}}
\newcommand{\gk}{g(\vk)}
\newcommand{\gzero}{g_0}
\newcommand{\geff}{g_\mathrm{eff}}
\newcommand{\Nph}{N_0}
\newcommand{\Ep}{\mathcal{E}}

\begin{document}

\title{Ultrafast All-Optical Switching via a Supersolid Phase Transition of Light}

\author{J.~L.~Figueiredo}
\affiliation{GoLP/Instituto de Plasmas e Fus\~{a}o Nuclear,
  Instituto Superior T\'{e}cnico, Universidade de Lisboa,
  1049-001 Lisboa, Portugal}

\author{J.~T.~Mendon\c{c}a}
\affiliation{GoLP/Instituto de Plasmas e Fus\~{a}o Nuclear,
  Instituto Superior T\'{e}cnico, Universidade de Lisboa,
  1049-001 Lisboa, Portugal}

\author{H.~Ter\c{c}as}
\affiliation{Instituto Superior de Engenharia de Lisboa,
  Instituto Polit\'{e}cnico de Lisboa,
  Rua Conselheiro Em\'{i}dio Navarro, 1959-007 Lisboa, Portugal}
\affiliation{GoLP/Instituto de Plasmas e Fus\~{a}o Nuclear,
  Instituto Superior T\'{e}cnico, Universidade de Lisboa,
  1049-001 Lisboa, Portugal}

\begin{abstract}
We propose ultrafast all-optical switching exploiting the bistability between a spatially uniform photon superfluid and a spontaneously ordered supersolid in a driven-dissipative microcavity. The key ingredient is a tunable nonlocal photon--photon interaction engineered by embedding a high-mobility two-dimensional electron gas (2DEG) inside the cavity. A drift current displaces the Fermi disk, imparting a negative region to the Lindhard interaction kernel at finite wavevectors and triggering a roton instability. The resulting bistable $S$-curve supports a write--hold--erase protocol in which short optical pulses toggle the system between branches with a switching contrast of order 120~dB in our simulations. The hysteretic ON state persists under a constant sub-threshold drive after the write pulse is removed, realizing an all-optical bistable memory. Since the photon field couples additively to each embedded quantum well, stacking layers with distinct drift angles allows the roton profile to be engineered with higher-order symmetries, imprinting richer spatial order on the supersolid and enabling nonbinary generalizations of the switch. Operating in the ultrafast, sub-fJ regime, this platform outperforms most existing all-optical switches in contrast and reconfigurability.
\end{abstract}

\maketitle

\textit{Introduction} --- Photonic information processing relies on optical switches that control the transmission or routing of light using light
itself. Such devices constitute the fundamental nonlinear
building blocks of optical logic, signal regeneration, and
reconfigurable photonic circuits, and are widely regarded as
essential components for future optical and neuromorphic
computing architectures \cite{Miller2010,Miller2017,Shen2017}. Modern information processing demands all-optical switches to be
simultaneously fast, energy-efficient, and
reconfigurable -- a combination that no existing platform
delivers~\cite{Nozaki2010,Caulfield2010}.
Kerr-based microcavity switches achieve low switching energy
($\lesssim1$~fJ) but are volatile, require active hold power, and
offer only moderate extinction ratios ($\sim20$--$30$~dB)~\cite{Nozaki2010,Almeida2004,Tanabe2005}.
Phase-change materials provide non-volatility but operate in the
nanosecond regime, are energy-intensive per write cycle, and cannot
be reconfigured without replacing the active medium~\cite{Wuttig2007,Rios2015}.
Semiconductor optical amplifiers (SOAs) deliver high contrast but
require continuous-wave (CW) gain and consume milliwatt-scale hold
power~\cite{Mork1999}.
Polariton switches exploit ultrafast exciton-photon coupling yet
remain volatile and deliver modest contrast ($\sim10$--$20$~dB)~\cite{Ballarini2013,Amo2009,Sanvitto2016}.
A fundamentally different approach is needed, one in which
the switching contrast originates from a sharp \emph{collective}
state change rather than from a weak local nonlinearity.

Driven-dissipative quantum fluids of light have emerged as a rich
platform for nonlinear photonics~\cite{Carusotto2013,Lagoudakis2008}.
Photon Bose--Einstein condensates (BEC) in dye-filled
microcavities~\cite{Klaers2010} and in semiconductor
microcavities~\cite{Schofield2024,Pieczarka2024} demonstrate that
macroscopic coherent photon condensates can be sustained at room
temperature, while exciton-polariton condensates achieve ultrafast
quantum-fluid dynamics~\cite{Weisbuch1992,Kasprzak2006,Amo2009}.
In a previous work~\cite{Figueiredo2025} we showed that embedding
a biased two-dimensional electron gas (2DEG) inside such a cavity
generates a nonlocal, momentum-selective photon--photon interaction
with negative regions at
finite wavevectors, driving a superfluid--to--supersolid transition
of the intracavity photon field.
The supersolid is characterized by simultaneous long-range density
order and global phase coherence, constituting a photonic analog of the supersolid phase previously observed
in ultracold atomic gases with dipolar
interactions~\cite{Tanzi2019,Guo2019,Norcia2021,Bottcher2019} and
in spin-orbit-coupled BECs~\cite{Leonard2017,Li2017}.

In this Letter we exploit the coexistence of two stable branches in
the driven-dissipative relation described in Ref.~\cite{Figueiredo2025} as a switching primitive. The OFF state is the spatially uniform superfluid, emitting
predominantly at $k\simeq0$, while the ON state is the supersolid that redistributes intensity into
Bragg-diffracted channels at $|\vk|=\kst$.
Because the transition is driven by a collective phase instability,
the contrast between branches reflects the macroscopic photon
redistribution, here reaching $124$~dB according to our simulations.
Crucially, the two bistable branches here are qualitatively distinct spatial phases rather than uniform states of differing density as in conventional driven-dissipative bistability~\cite{Gibbs1985,FossFeig2017}.
Once written, the ON state is retained under the same CW hold pump
without reapplying the write pulse. Moreover, the lattice orientation is
electrically reconfigurable, making this the first photonic switch
to simultaneously achieve high contrast, bistable memory, and
multiport programmability.

\textit{Model} --- Let us consider a semiconductor microcavity sustaining a two-dimensional massive photon gas under sufficiently strong longitudinal confinement, as described in Ref.~\cite{Figueiredo2025}. In convenient units [see the Supplemental Material (SM)~\cite{SM} for more details], the intracavity field envelope $\Ep(\bvr,t)$ in the rotating frame
obeys the driven-dissipative nonlocal Gross--Pitaevskii (GP) equation~\cite{Figueiredo2025}:
\begin{equation}
  i\frac{\partial}{\partial t} \Ep
  = \left[-\tfrac{1}{2}\nabla^2 \!-\! i\kappa \!-\! \Omega_p\right]\Ep
  + \alpha(g*|\Ep|^2)\Ep
  + \Gamma(t),
  \label{eq:GP}
\end{equation}
where the nonlinear potential $g*|\Ep|^2 \equiv \int d^2r'\,g(\bvr{-}\bvr')|\Ep(\bvr')|^2$ represents nonlocal photon--photon interactions, $\kappa$ is the cavity-loss rate, $\Omega_p$ the pump frequency,
$\alpha$ the nonlinear coupling strength, and $\Gamma(t)$ the coherent
pump amplitude. Note that the last three quantities play distinct roles: $\Gamma(t)$ is the external coherent drive injected through the cavity mirror and is the only source term; $\alpha$ fixes the overall magnitude of the 2DEG-mediated photon--photon interaction; and $\gk$ carries its momentum dependence. No light--matter coupling constant appears explicitly because the electronic degrees of freedom have been integrated out, leaving $\alpha \gk$ as the induced photon nonlinearity.
This driven-dissipative GP framework generalizes the equilibrium
Gross-Pitaevskii theory~\cite{Pitaevskii2003,Dalfovo1999} to
open quantum systems and has been extensively validated for polariton
condensates~\cite{Carusotto2013,Wouters2007}.

\begin{figure}[t!]
  \centering
  \includegraphics[scale=0.52]{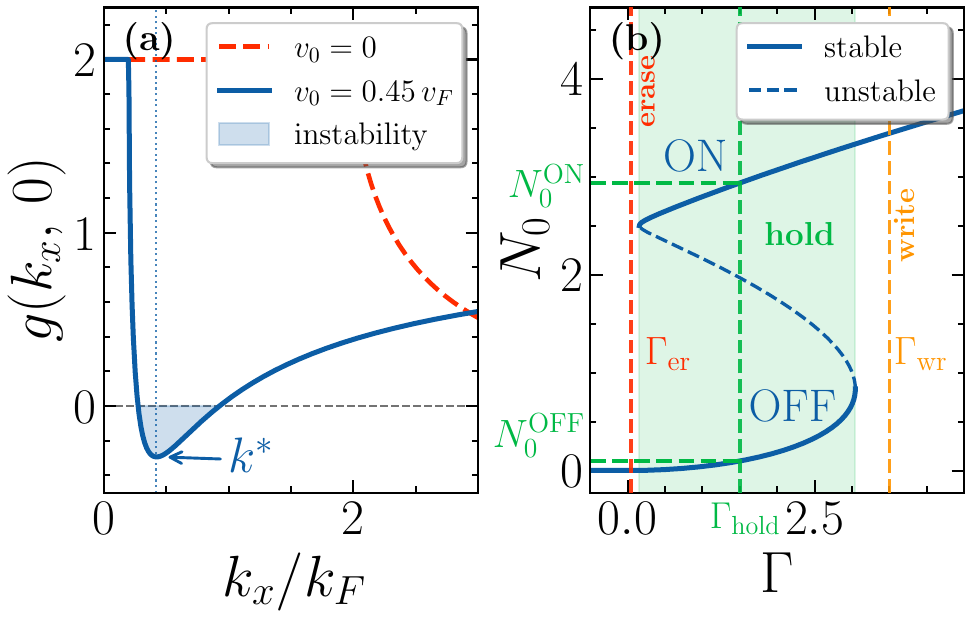}
  \caption{%
	Interaction kernel and bistability.
    (a)~Interaction ernel $g(k_x,0)$ (in units of $\frac{e^4}{2\pi \hbar \epsilon_0 d_0  \omega_0^3}$) along the drift direction
    for $v_0=0$ (dashed, repulsive and flat) and
    $v_0=0.45\,\vF$ (solid).
    The shaded area marks the negative-$g$ instability band;
    the arrow indicates the roton minimum at $\kst \simeq 0.42\,\kF$.
    (b)~Steady-state photon density $\Nph$ vs.\ pump amplitude $\Gamma$
    ($S$-curve) for fixed $\Omega_p = 5$, $\alpha = 1$, $\kappa = 0.1$. All quantities are in dimensionless units. The shaded band is the bistable window $\Gamma\in(\Gamma_\text{min},\Gamma_\text{max})$.
    Dashed vertical lines marked at $\Gamma_{\rm er}$, $\Gamma_{\rm hold}$, and $\Gamma_{\rm wr}$ corresponding to the erase, hold and write pulse amplitudes; dashed horizontal lines marked at $N_0^\text{OFF}$ (label shifted up for visibility) and $N_0^\text{ON}$ corresponding to the OFF and ON states, respectively. The surrounding white zones are the single-valued erase and write regions, respectively.
  }
  \label{fig:kernel}
\end{figure}

The nonlocal kernel $\gk$, defined as the Fourier transform of $g(\mathbf r)$ in Eq.~\eqref{eq:GP}, is related to the static Lindhard
polarizability of the 2DEG~\cite{Figueiredo2025} by
\begin{equation}
  g(\vk) = - G_0
\,\chi_0\!\bigl(\vk,\,\omega = -\vk\cdot\vv_0\bigr),
  \label{eq:gk}
\end{equation}
where $G_0= e^4\hbar/(m^2\epsilon_0 d_0 \omega_0^3)$, $m$ being the effective electron mass, $d_0$ the cavity distance, and $\vv_0 = \hbar \vk_0/m$ is the drift velocity of the 2DEG externally imposed. Physically, the intracavity photon field couples to virtual
intraband particle--hole excitations of the 2DEG Fermi sea (an electron promoted above the Fermi level, leaving an empty state below it)
generating a nonlocal photon--photon interaction mediated by
charge-density fluctuations.
For $v_0 = 0$, $\gk$ is isotropic and everywhere positive, yielding only
repulsive interactions and superfluid behavior
[Fig.~\ref{fig:kernel}(a), dashed curve], while a drift $v_0 = 0.45\,\vF$ introduces a Doppler shift
$\omega = -\vk\cdot\vv_0$ into the susceptibility. Consequently, photon modes
satisfying $|\hbar \vk\cdot\vv_0| < \varepsilon_F$, with $ \varepsilon_F$ the Fermi energy, acquire an imaginary part that, through the Kramers--Kronig
relations, drives the real part negative near $\kst \simeq 0.42\,\kF$,
with minimum $g(\kst) \simeq -0.29$ in proper units
(see Fig.~\ref{fig:kernel}(a); all normalizing units can be found in~\cite{SM}).
This attractive channel seeds a roton instability with spatial period
$a = 2\pi/\kst \simeq 15\,\kF^{-1}$, which is approximately $5\,\mu$m for the GaAs reference design specified below.

Multiport operation exploits the additive structure of the
Lindhard functions.
When $N_\mathrm{QW}$ quantum-well layers, each carrying a drift current at
angle $\phi_j$ with respect to the $x$-axis, are stacked in the same cavity
with sufficient separation so that interlayer electron--electron interactions
can be neglected, the intracavity photon field interacts with each layer
independently. The total effective photon--photon kernel is therefore
\begin{equation}
  \geff(\vk) = \sum_{j=1}^{N_\text{QW}} g_j(\vk) - (N_\text{QW}-1) g_0 ,
  \label{eq:geff} 
\end{equation}
where $g_j(\vk)$ is the Lindhard kernel of layer $j$ and the last term ensures that $g_\text{eff}(\mathbf k = 0) = g_0$.
By choosing the set of drift directions $\{\phi_j\}$ one sculpts
$\geff$ to display roton minima at any desired constellation of
wavevectors, enabling selection of the supersolid lattice symmetry entirely by electrical gate control.

\begin{figure}[t!]
  \centering
  \hspace{-0.2cm}
  \includegraphics[scale=0.48]{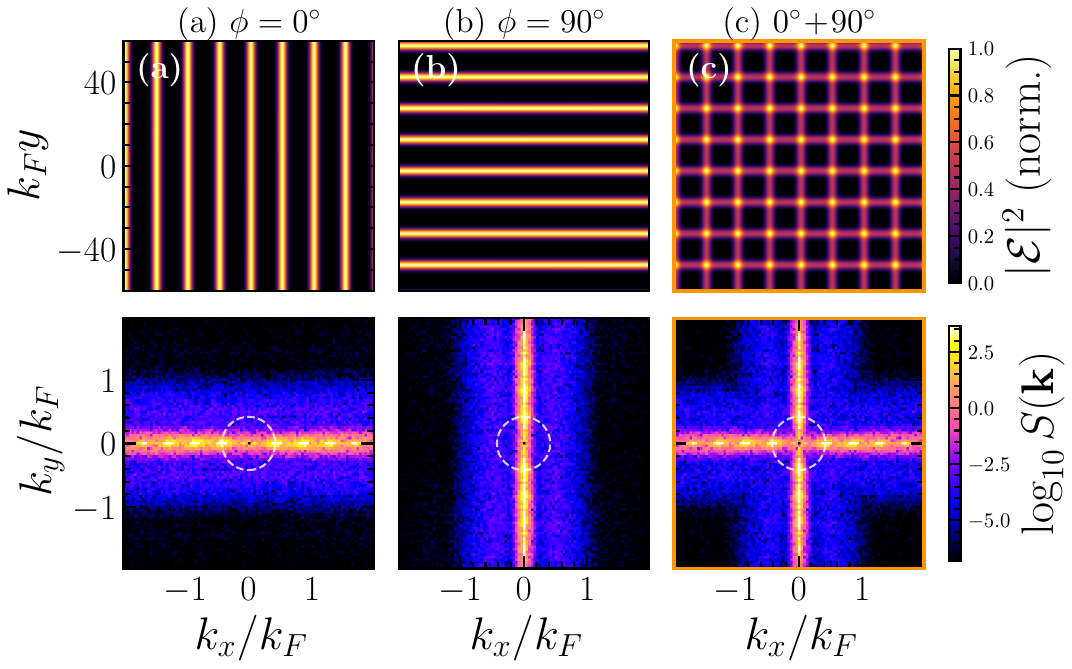}
  \caption{Reconfigurable supersolid lattice.
    Top row: steady-state density $|\Ep(\bvr)|^2$.
    Bottom row: structure factor $S(\vk)$ in logarithmic scale;
    dashed circle marks $|\vk|=\kst$.
    (a)~Single layer, $\phi=0^\circ$: vertical stripes, Bragg peaks
    at $(\pm\kst,0)$.
    (b)~Single layer, $\phi=90^\circ$: horizontal stripes, Bragg peaks
    at $(0,\pm\kst)$.
    (c)~Two layers ($\phi_1=0^\circ$, $\phi_2=90^\circ$, orange frame):
    single intracavity field with effective kernel $\geff=g_1+g_2$;
    simultaneous roton minima at $(\pm\kst,0)$ and $(0,\pm\kst)$
    seed a square supersolid with fourfold Bragg symmetry.
    All configurations are switched by reprogramming the dc gate bias, and no optical realignment is required.
    Parameters (in the dimensionless units of the SM~\cite{SM}, i.e.\ energies in
    units of $\hbar^2\kF^2/M$ and lengths in units of $\kF^{-1}$):
    $\Omega_p = 5$, $\Gamma = 1.5$, $\alpha = 1$,
    $\kappa = 0.1$, $v_0 = 0.45\,\vF$.
  }
  \label{fig:reconfig}
\end{figure}

For two orthogonal layers ($\phi_1=0^\circ$, $\phi_2=90^\circ$),
$\geff(\vk)$ acquires simultaneous roton minima
at $(\pm\kst,0)$ and $(0,\pm\kst)$, seeding a square supersolid
with fourfold Bragg symmetry [see Fig.~\ref{fig:reconfig}(c)].
More generally, $N_\text{QW}$ layers with angles $\phi_j = j\pi/N_\text{QW}$
generate $2N_\text{QW}$--fold roton minima, producing a honeycomb supersolid 
for $N_\text{QW}=3$, or quasicrystalline patterns for $N_\text{QW}\geq 4$, 
with octagonal symmetry at $N_\text{QW}=4$.\\

\textit{Bistability and Switching Protocol} --- For a spatially uniform steady state
$\Ep(\bvr,t) = \mathcal E_0 e^{-i\Omega_p t}$ and defining $g_0 \equiv g(\mathbf k = 0)$,
the superfluid density $\Nph = |\mathcal E_0|^2$ satisfies the cubic
\begin{equation}
  \alpha^2 g_0^2 \Nph^3 - 2\Omega_p\alpha g_0 \Nph^2
  + \bigl(\Omega_p^2 + \kappa^2\bigr)\Nph = \Gamma^2,
  \label{eq:cubic}
\end{equation}
which has either one or three real roots depending on the parameters, thereby defining a bistable region~\cite{Figueiredo2025}.
Equation~\eqref{eq:cubic} is structurally equivalent to the Drummond-Walls cubic for a driven-dissipative Kerr cavity~\cite{Drummond1980}, here with an upper-branch solution that lies above the roton instability threshold so the ON state is a supersolid rather than a mere denser uniform superfluid.
Figure~\ref{fig:kernel}(b) shows the resulting $S$-curve for
$\Omega_p = 5$, $\alpha = 1$, $\kappa = 0.1$.
The lower (upper) stable branch carries the OFF (ON) state, with the bistable
window spanning $\Gamma \in (\Gamma_\mathrm{min},\Gamma_\mathrm{max}) \simeq (0.16, 3.04)$. For a fixed pump value $\Gamma_\text{hold} = 1.5$ located at the center of the bistable region, Eq.~\eqref{eq:cubic} yields $ N_0^\mathrm{OFF} \simeq 0.097$ on the lower branch
(spatially uniform superfluid) and $N_0^\mathrm{ON} \simeq 2.94$ on the upper branch, lying above the roton instability threshold as indicated in Fig.~\ref{fig:kernel}(b). Unlike prior work on optical bistability in polariton microcavities~\cite{Gippius2007,Magnusson2011}, where both bistable branches are spatially uniform condensates differing only in density or polarisation, here the ON branch spontaneously orders into a supersolid, which leads to large switching contrasts. Additionally, the value of $g_0$ is the zero-momentum Lindhard polarizability of the drifting 2DEG, which depends on the drift velocity $v_0$ and is electrically tunable. This allows the bistable window $[\Gamma_{\min}, \Gamma_{\max}]$ to be programmed via a gate bias, without any optical reconfiguration.

\begin{figure}[t!]
  \centering
    \includegraphics[scale=0.55]{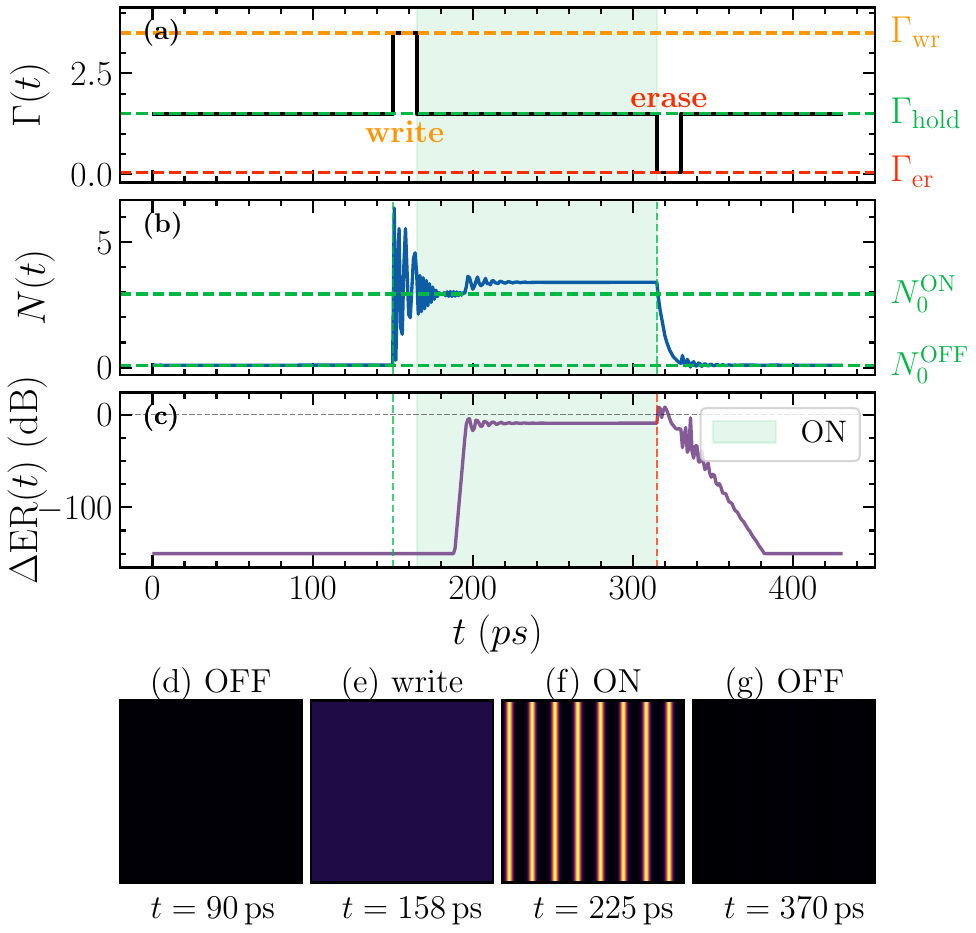}
   \caption{
Switching dynamics. Time is shown in picoseconds and all other quantities are in the dimensionless units of the SM~\cite{SM}. On this scale the photon lifetime is $\kappa^{-1} \simeq 10$~ps and the full write--hold--erase cycle spans ${\simeq}\,400$~ps. The shaded region is the ON-state hold time.
    (a)~Pump protocol $\Gamma(t)$: CW hold drive ($\Gamma_\text{hold}=1.5$,
    dotted) interrupted by a write pulse ($\Gamma_\text{wr} = 3.5$ and duration $15\,$ps). Later an erase pulse is applied ($\Gamma_\text{er}=0.05$, same duration). 
    (b)~Mean intracavity photon density $N(t) = \int d^2r |\mathcal E(r)|^2/A$ with $A$ the total area, equaling $N_0^\mathrm{OFF}$ ($N_0^\mathrm{ON}$) on the lower (upper) branch of the $S$-curve in the plateaus of Fig.~\ref{fig:kernel}(b), corresponding to a total number of photons of approximately 2200 (74000) for a simulation box with length $L=150\,\kF^{-1}$.
    (c)~Extinction ratio
    $\mathrm{ER}(t) = 10\log_{10}(I_\mathrm{Bragg}/I_\mathrm{through})$;
    the switching contrast is $\Delta\mathrm{ER} = 124$~dB.
    (d)--(g) Density snapshots $|\Ep(\bvr)|^2$ at representative times:
    (d)~OFF (featureless superfluid);
    (e)~during write pulse (transient);
    (f)~ON (stripe supersolid, wavenumber $\kst$);
    (g)~OFF.
  }
  \label{fig:switching}
\end{figure}

The roton instability condition for the upper-branch state reads
$\eta(\kst)[\eta(\kst)+2\alpha\Nph g(\kst)] < -\kappa^2$,
where $\eta(\vk) = k^2/2 - \Omega_p + \alpha\Nph\gzero$~\cite{SM}.
For the parameters above, the left-hand side evaluates to $-0.596$ 
at $\kst \simeq 0.42\,\kF$ and $N_0 = \Nph^\mathrm{ON}$, confirming that the ON branch is supersolid.
On the lower branch at $N_0 = \Nph^\mathrm{OFF}$ the same expression is positive,
confirming the OFF branch is a stable, structureless superfluid.
The characteristic roton growth rate on the upper branch follows from
the driven-dissipative Bogoliubov dispersion relation~\cite{Figueiredo2025,SM}:
\begin{equation}
    \Gamma_\mathrm{rot}  = \sqrt{\eta(\kst)
  \bigl[\eta(\kst)+2\alpha\Nph g(\kst)\bigr] - \kappa^2}
,
  \label{eq:growth}
\end{equation}
which at the conditions considered here becomes $\Gamma_\mathrm{rot} \sim \kappa$, leading to a crystallization time $\tau_e = \Gamma_\mathrm{rot}^{-1} \sim \kappa^{-1}$.
The observed switching time $\tau_\mathrm{sw} \sim 50$ (in dimensionless units, see Fig.~\ref{fig:switching} for numerical values) is much longer than $\tau_e$ because the supersolid does not start from a macroscopic seed. Contrarily, it must grow from
quantum noise.
Starting at $\delta\Nph/\Nph \sim 10^{-3}$, the roton mode must be
amplified by a factor of $10^{3}$ before the density pattern becomes
macroscopic, which requires $\ln(10^{3})\approx 7$ successive 
crystallization periods. The remaining delay arises from nonlinear saturation as the system locks
into the supersolid steady state.
This picture is directly confirmed by simulations that track
$S(\kst,t)$ starting from the upper-branch homogeneous state plus noise. We observe that
$S(\kst)$ grows exponentially with rate $2\Gamma_\text{rot}\approx1.52\,t_0^{-1}$,
in quantitative agreement with Eq.~\eqref{eq:growth}, before saturating
at $S^\mathrm{ss}\approx12$ -- see Sec.~VI of the SM~\cite{SM} for additional discussion.
Notably, erasing is faster than writing as the erase pulse removes
the drive entirely ($\Gamma_\mathrm{er}=0.05$), so the field
decays on the loss timescale $\kappa^{-1}$ without requiring
any instability to develop.

The switching protocol is described in Fig.~\ref{fig:switching}(a) and exploits these two stable branches.
A CW hold drive at $\Gamma_\mathrm{hold} = 1.5$
keeps the system inside the bistable window on the lower branch (OFF).
The hold point lies well within the window -- a factor of $\sim9$
above the lower fold and a factor of $\sim2$ below the upper
fold -- providing a wide tolerance against pump fluctuations.
A \emph{write pulse}, i.e., a brief increase to $\Gamma_\mathrm{wr}
= 3.5$, above the upper saddle-node, pushes the system into the
upper-branch.
After the pulse, the field relaxes to the ON state and remains there
under the hold drive as the switch is latched with no power cost
beyond the threshold.
Then, an \emph{erase pulse} consisting of a brief reduction to $\Gamma_\mathrm{er}
= 0.05$, located below the lower saddle-node at $\Gamma = 0.16$, resets
the system to OFF within a time of order $\sim\kappa^{-1}$.

\textit{Switching dynamical simulations} ---
Figure~\ref{fig:switching} shows a complete \emph{write--hold--erase} cycle
obtained by time-integrating Eq.~\eqref{eq:GP} numerically as described in the SM~\cite{SM}. Starting from the OFF state ($\Nph \simeq 0.097$), the write pulse
drives the field to the ON state within
$\tau_\mathrm{sw} \simeq 50\,t_0$, i.e.\ ${\simeq}\,50$~ps for the GaAs reference
design ($t_0 \simeq 1$~ps) used consistently throughout this Letter.
The whole cycle can be read directly off the $S$-curve of
Fig.~\ref{fig:kernel}(b), whose vertical and horizontal markers carry the same
labels as Fig.~\ref{fig:switching}.
The hold drive $\Gamma_\mathrm{hold}$ fixes the operating point on the lower
branch at $N(t) = N_0^\mathrm{OFF}\simeq0.097$, while the write pulse
$\Gamma_\mathrm{wr}>\Gamma_\mathrm{max}$ carries it past the upper fold and
commits it to the upper branch, where it remains at $N_0^\mathrm{ON}\simeq2.94$
after $\Gamma$ returns to $\Gamma_\mathrm{hold}$. The erase pulse
$\Gamma_\mathrm{er}<\Gamma_\mathrm{min}$ returns it below the lower fold to
$N_0^\mathrm{OFF}$.
The intracavity density $N(t)$ in Fig.~\ref{fig:switching}(b) is the full photon density tracing the
hysteresis loop of Fig.~\ref{fig:kernel}(b).
The ON plateau settles slightly above the uniform reference $N_0^\mathrm{ON}$
because the physical ON state is the modulated supersolid rather than the
uniform upper-branch fixed point of Eq.~\eqref{eq:cubic} and displays a small excess density due to the nonlinear saturation of the roton instability.
After the pulse, the ON state is self-sustained by the hold drive and
the density field spontaneously organizes into sharp spatial stripes
with wavenumber $\kst \simeq 0.42\,\kF$ [Fig.~\ref{fig:switching}(f)],
while the OFF state is dark and featureless [Fig.~\ref{fig:switching}(d)].

The extinction ratio
$\mathrm{ER}(t) = 10\log_{10}(I_\mathrm{Bragg}/I_\mathrm{through})$,
where $I_\mathrm{Bragg}$ ($I_\mathrm{through}$) is the spectral
intensity near $|\vk| = \kst$ ($|\vk| \simeq 0$)~\cite{SM}, rises from
$\mathrm{ER}^\mathrm{OFF} \simeq -147$~dB to
$\mathrm{ER}^\mathrm{ON} \simeq -23$~dB
[Fig.~\ref{fig:switching}(c)]. The switching contrast $\Delta\mathrm{ER} = 124$~dB is a direct
consequence of the macroscopic photon redistribution associated with
the superfluid--to--supersolid phase transition. While the OFF-state
structure factor $S(\kst)$ ideally vanishes, in real devices it is limited by thermal fluctuations and noise populating the Bragg modes. The value quoted here is therefore the contrast of a noise-free simulation and should be read as an upper bound. In a real device the measurable contrast is set by the detector noise floor and by residual disorder scattering into the Bragg channels, and we expect it to remain well above the $20$--$30$~dB typical of resonance-based switches. Contrarily, in the ON state the finite supersolid order parameter concentrates most of the intracavity power into the
$|\vk| = \kst$ Bragg modes, with the remaining power in the
forward channel.
The asymmetric ER values reflect the
threshold nature of the phase transition. The OFF-to-ON jump is not
a continuous shift but a discontinuous change in the structural
order of the photon field driven by quenching $\Gamma(t)$.
During the write pulse, the average photon density $N(t)$ 
transiently overshoots its asymptotic ON value as depicted in Fig.~\ref{fig:switching}(b), which is a signature of roton
amplification preceding nonlinear saturation via mode competition.
The erase pulse restores $\mathrm{ER}$ to its baseline
within $\sim\kappa^{-1}$, completing the hysteretic cycle. 

\begin{table}[b!]
\label{tab:comparison}
\begin{ruledtabular}
\begin{tabular}{lcccc}
Platform & $\Delta$ER (dB) & $\tau_\mathrm{sw}$ & BM & RC \\
\hline
Supersolid (this work)                    & $\lesssim$124 & $\sim$50 ps  & \checkmark & \checkmark \\
Polariton bistable~\cite{Ballarini2013}   & 10--20       & 1--10 ps        & $\times$   & $\times$   \\
Si micro-ring~\cite{Nozaki2010}           & 20--30       & 10--100 ps      & $\times$   & $\times$   \\
Phase-change (GST)~\cite{Wuttig2007}      & 20--30       & $\gtrsim$1 ns   & \checkmark & $\times$   \\
III-V SOA~\cite{Ju} & 5--10 & $\lesssim 1$~ps & $\times$ & $\times$ \\
Atom/EIT~\cite{Chang2014}                 & 10--20       & $\mu$s--ms      & $\times$   & $\times$   \\
\end{tabular}
\end{ruledtabular}
\caption{%
  Comparison of all-optical switching platforms.
  $\Delta\mathrm{ER}$: ON/OFF contrast in the designated output
  channel. $\tau_\mathrm{sw}$: switching time. BM: bistable memory
  (ON state retained under CW hold drive after write pulse). RC: in-situ reconfigurable
  routing. References indicate representative experimental demonstrations.
}
\label{tab:comparison}
\end{table}

\textit{Supersolid switch performance} --- Table~\ref{tab:comparison} benchmarks these results against leading
all-optical switch platforms, showing that the supersolid switch combines high contrast, bistable memory and in-situ reconfigurability in a way that no single existing platform offers simultaneously: (i) The contrast $\Delta\mathrm{ER} = 124$~dB is set by the
macroscopic redistribution of the photon field between structural
phases and is therefore
decoupled from the trade-off between speed and modulation depth
that limits resonance-based devices.
(ii) The switch memory is hysteretic: the ON (supersolid) state
is a locally stable fixed point of the GP equation for hold pumps
sufficiently above the lower fold, so after the write pulse the system remains on the
upper branch under the same CW hold drive.
This is in contrast to Kerr micro-cavity switches, which return to
the OFF state as soon as the write pulse ends, even if a hold pump
is maintained.
(iii) The drift direction is set by a weak dc bias ($\sim$1--10~mV
across a $\sim$10--30~$\mu$m quantum-well), tunable on nanosecond
timescales without disrupting the intracavity field.
(iv) The device operates in the \emph{weak-coupling} (photon BEC)
regime~\cite{Schofield2024,Pieczarka2024,Klaers2010}, so exciton-polariton strong
coupling is not required~\cite{Weisbuch1992}, removing the stringent
spectral matching constraints of polariton platforms and simplifying
fabrication.

Noise robustness follows from the large bistable separation.
Thermal and quantum fluctuations in the intracavity photon density
have amplitude $\delta N  \sim \sqrt{ N } \sim 0.3$ for the
lower-branch equilibrium, while the gap between the two steady-state values is $\Delta N  \simeq 3.2$, more than an order of magnitude larger.
Spontaneous switching is therefore exponentially suppressed.
The main practical concern is pump-amplitude noise $\delta\Gamma$, since 
$\Gamma_\mathrm{hold}$ drifting sufficiently close to  the saddle point may promote noise-driven inter-branch transitions. Our choice $\Gamma_\mathrm{hold} = 1.5$ sits roughly in the center
of the bistable window, providing a margin of $\simeq
9$ times from the lower fold and $\simeq 2$ times from the upper fold.
Further parameter scans are presented in Sec.~VIII of~\cite{SM} and confirm that switching is robust for
$\Gamma_\mathrm{wr}\gtrsim 3.0$  and that the ON state persists across the range $\Gamma_\mathrm{hold}\in[1.0,\,2.8]$ covering a factor of ${\simeq}\,5$ in hold-drive amplitude above the lower bistable limit.

The stripe supersolid spontaneously breaks continuous translational
symmetry along the drift axis, selecting a stripe phase $\phi_0$ from
a degenerate manifold $[0,\,2\pi/\kst)$.
This gives rise to a gapless Goldstone mode -- the \emph{phason}~\cite{fishman1996goldstone,kaplan2025optically,gao2022symmetry} -- that
governs slow diffusion of the stripe position under noise.
For switching, the phason is entirely benign since the Bragg power at
$|\vk| = \kst$ is insensitive to $\phi_0$ and only the real-space stripe position drifts, as confirmed by our simulations. 
This contrasts favorably with phase-change materials, where repeated
write-erase cycles reset the crystalline domain structure
stochastically, degrading device reproducibility.
The phason also implies that two supersolid domains can merge
adiabatically if their stripe phases are compatible, facilitating
the cascaded operation of coupled-cavity arrays.

After each erase event, the field decays to the OFF state on the
loss timescale $\kappa^{-1}$, setting a maximum write-erase
repetition rate $f_\mathrm{rep} \sim \kappa$.
For $\kappa \sim 0.1$~ps$^{-1}$ this gives $f_\mathrm{rep} \sim
100$~GHz, limited by the photon lifetime rather than by
phase-transition kinetics.
This is orders of magnitude faster than PCM devices
($\lesssim 1$~GHz, limited by material crystallization dynamics~\cite{Wuttig2007})
and competitive with the fastest Kerr resonator switches~\cite{Nozaki2010}.
The bandwidth could be further increased by raising $\kappa$
(shorter cavity lifetime) at the cost of a higher hold-drive
threshold, a trade-off governed directly by the bistable window
boundaries in Eq.~\eqref{eq:cubic}.
 
\textit{Experimental set-up} --- The target experimental system is a GaAs distributed-Bragg-reflector
(DBR) microcavity with embedded GaAs/Al$_{x}$Ga$_{1-x}$As quantum
wells as considered in Ref.~\cite{Figueiredo2025}.
A 2DEG is formed by modulation doping a nearby AlGaAs layer to carrier
density $n_e \sim 10^{10}$--$10^{11}$~cm$^{-2}$~\cite{Stormer1983,Pfeiffer1989} and
the drift current is set by ohmic contacts on opposite sides of the QW.
The switch state is read by far-field detection at the Bragg angle
$\theta_B = \sin^{-1}(\kst/p_\mathrm{ph})$ relative to the
cavity axis, easily separated from the forward output by standard
spatial filtering.
Existing 2DEG--microcavity experiments operate at zero drift current, where $\gk$ is isotropic and positive definite and no roton minimum exists. The finite drift $v_0 \simeq 0.45\,\vF$ that opens the attractive channel at $\kst$ is precisely the ingredient that has not yet been implemented.
Moreover, the effective photon mass $M \sim 10^{-5}\,m_e$ and
$\kF \sim 1$--$10\,\mu\mathrm{m}^{-1}$ place the dimensionless time
unit $t_0 = M/(\hbar\kF^2)$ in the range $0.1$--$10$~ps, so our
switching time $\tau_\mathrm{sw} \simeq 50\,t_0$ spans $\sim 5$--$500$~ps across
this window. All physical numbers quoted in this Letter refer to the reference
design $\kF \simeq 3\,\mu\mathrm{m}^{-1}$, for which $t_0 \simeq 1$~ps,
$\tau_\mathrm{sw} \simeq 50$~ps, the photon lifetime is
$\kappa^{-1} = 10\,t_0 \simeq 10$~ps and the supersolid period is
$a \simeq 5\,\mu$m.
For a photon lifetime $\tau_c \sim 10$~ps
($\kappa \sim 0.1$~ps$^{-1}$) and intracavity photon number
$N \sim 10^3$, the write-pulse energy is
$\Delta E_\mathrm{sw} \sim \omega_0\,\Delta N \,\tau_\mathrm{pulse}
\sim 0.1$--$1$~fJ, comparable to state-of-the-art microresonator
switches~\cite{Nozaki2010,Reed2010}.
The minimum pulse duration needed to commit the system to the ON state is
$t^*_\mathrm{wr} \approx 0.4\,t_0 \approx 0.04\,\kappa^{-1}$, only a
twentieth of the photon lifetime, as established by a write-pulse duration
scan at fixed $\Gamma_\mathrm{wr} = 3.5$ through additional simulations that can be found in Sec.~VII of~\cite{SM}.
Beyond GaAs, transition-metal dichalcogenide (TMD) heterostructures
(e.g.\ MoSe$_2$/WSe$_2$) host high-mobility 2DEGs at room temperature
with $\kF \sim 10$~nm$^{-1}$, compressing the supersolid period to
$a \sim 1$~nm and pushing operation toward near-UV wavelengths,
while graphene offers room-temperature ballistic transport
compatible with photonic waveguide integration~\cite{Basov2016,Low2017}.

\textit{Reconfigurability}---
Figure~\ref{fig:reconfig} demonstrates in-situ reconfigurability
by changing the drift direction $\phi$.
For $\phi = 0^\circ$ (drift along $x$), the roton instability is
triggered along $k_x$, forming vertical density stripes with
Bragg peaks at $(\pm\kst, 0)$ [Fig.~\ref{fig:reconfig}(a)].
Rotating the drift by $90^\circ$ transfers the instability to $k_y$,
producing horizontal stripes with peaks at $(0, \pm\kst)$
[Fig.~\ref{fig:reconfig}(b)].
In both cases, changing $\phi$ reroutes the diffracted output between
orthogonal far-field ports without any optical realignment, and lattice direction is set entirely by a dc gate bias
applied to the 2DEG. \par 

By stacking two layers with orthogonal drift directions
($\phi_1=0^\circ$, $\phi_2=90^\circ$), the effective kernel
$\geff = g_1 + g_2 - g_0$ acquires simultaneous roton minima at
$(\pm\kst,0)$ and $(0,\pm\kst)$, driving the photon field into a
two-dimensional square supersolid with four Bragg peaks
[Fig.~\ref{fig:reconfig}(c)].
This extends the binary OFF/ON switch to a \emph{three-state} device:
the OFF state is the uniform superfluid; ON$_1$ (stripe, $\phi_2=0^\circ$)
carries Bragg peaks at $(\pm\kst,0)$ only; ON$_2$ (square, $\phi_2=90^\circ$)
carries peaks at all four $(\pm\kst,0)$, $(0,\pm\kst)$ positions.
The desired ON state is selected by setting $\phi_2$ via a dc gate-bias
reversal \emph{before} the write pulse, while the pump amplitude protocol is
identical for both ON states.
In Sec.~IX of the SM~\cite{SM} we present a detailed protocol and 
additional simulations of the complete write--hold--erase cycle confirming that both ON
states are bistable, with the structure-factor ratio
$S_y/S_x \approx 0$ for ON$_1$ and $S_y/S_x \approx 1$ for ON$_2$,
uniquely identifying each state. Adding further layers increases the number of ON states with higher-symmetry
supersolid order, each accessible by the same bistable writting mechanism.

{\it Conclusions}--- We have demonstrated by simulations that a
driven-dissipative photon condensate coupled to a biased 2DEG
supports a hysteretic superfluid--supersolid phase transition that
functions as an ultrafast, bistable, reconfigurable all-optical
switch with contrast values that can surpass $100$~dB.
The switching mechanism is fundamentally different from both
resonance-shift (Kerr, EO) and absorption-saturation (SOA, polariton)
schemes. The contrast originates from a macroscopic
redistribution of photon momentum between structural phases,
bypassing all conventional material speed-energy trade-offs.
The device simultaneously provides logical state storage (via
hysteretic bistability), routing programmability (via drift angle),
and multiport operation (via Bragg diffraction), three capabilities
that have not previously been combined in a single photonic device.
Arrays of coupled cavities could further extend the platform to
optical logic based on domain walls and phase textures of supersolid
light~\cite{Ostermann2016}, or to photonic quantum simulation of interacting lattice
models~\cite{Carusotto2013,Hartmann2008,Tomadin2010}.

\begin{acknowledgments}
J.~L.~F.\ acknowledges Funda\c{c}\~{a}o para a Ci\^{e}ncia e a
Tecnologia (FCT-Portugal) through Grant No.~UI/BD/151557/2021.
\end{acknowledgments}

\bibliographystyle{apsrev4-1}
\bibliography{references}

\end{document}


\setcounter{secnumdepth}{3}

\title{Supplemental Material: ``Ultrafast All-Optical Switching via a Supersolid Phase Transition of Light''}

\author{J.~L.~Figueiredo}
\affiliation{GoLP/Instituto de Plasmas e Fus\~{a}o Nuclear,
  Instituto Superior T\'{e}cnico, Universidade de Lisboa,
  1049-001 Lisboa, Portugal}

\author{J.~T.~Mendon\c{c}a}
\affiliation{GoLP/Instituto de Plasmas e Fus\~{a}o Nuclear,
  Instituto Superior T\'{e}cnico, Universidade de Lisboa,
  1049-001 Lisboa, Portugal}

\author{H.~Ter\c{c}as}
\affiliation{Instituto Superior de Engenharia de Lisboa,
  Instituto Polit\'{e}cnico de Lisboa,
  Rua Conselheiro Em\'{i}dio Navarro, 1959-007 Lisboa, Portugal}
\affiliation{GoLP/Instituto de Plasmas e Fus\~{a}o Nuclear,
  Instituto Superior T\'{e}cnico, Universidade de Lisboa,
  1049-001 Lisboa, Portugal}

\maketitle

\tableofcontents

\section{Drive-dissipative Gross-Pitaevskii equation}
Considering the intracavity field $\bm{\mathcal E}(\mathbf r,t) = \mathcal E(\mathbf r,t) \mathbf e_z$ along the $z-$direction, the full Gross-Pitaevskii equation has been determined in Ref.~\cite{S9} to be 
\begin{align}
&i\hbar \frac{\partial}{\partial t}\mathcal E(\mathbf r,t) = \Big[-\frac{\hbar^2}{2M} \bm{\nabla}_\perp^2 + V_\text{trap}(\mathbf r) -i\hbar \kappa \Big]\mathcal E (\mathbf r,t) \nonumber \\
&+ \hbar \Gamma E_p(\mathbf r,t)  + \int d\mathbf r' \, g(\mathbf r-\mathbf r') \big| \mathcal E(\mathbf r',t)\big|^2 \mathcal E(\mathbf r,t), \label{GP}
\end{align}
describing cavity photons with effective mass $M$ and trap potential $V_\text{trap}(\mathbf r)$, both fixed by the cavity geometry~\cite{S10,S11}. Above, $\kappa$ and $\Gamma$ are the loss and pump rate, $E_p(\mathbf{r}) = P e^{-i\Omega_pt}$ the pump field, $P = |P|e^{i\theta }$, $\theta$ being the pump phase and $\Omega_p$ the pump frequency. Most crucially, $g(\mathbf r)$ reads as a medium-induced nonlinear interaction kernel between photons. 
To ease the notation and facilitate our calculations, we work in dimensionless units throughout the paper by rescaling physical variables to the values given in Tab.~\ref{tab_normalization}. In these units and after moving to the rotated frame $\mathcal E(\mathbf r,t) \rightarrow  \mathcal  E(\mathbf r,t) e^{-i\Omega_p t}$, Eq.~\eqref{GP} becomes
\begin{equation}
  i\frac{\partial}{\partial t} \mathcal E
  = \left[-\frac{1}{2}\nabla^2 \!-\! i\kappa \!-\! \Omega_p\right] \mathcal E
  + \alpha(g*|\mathcal E|^2)\mathcal E + \Gamma,
  \label{GP_normalized}
\end{equation}
where $g*|\mathcal E|^2 \equiv \int d^2r'\,g(\mathbf r - \mathbf r')|\mathcal E(\mathbf r')|^2$ as given in the main text. Above we neglected the trap potential which is approximately constant for InGaAs across many typical distances of $1/k_F$.

The dimensionless coupling constant $\alpha$ controlling the nonlinear 
interaction strength is
\begin{equation}
    \alpha = \frac{e^4 |P|^2}{2\pi\hbar^3 k_F^2\epsilon_0 d_0\omega_0^3},
    \label{eq:alpha}
\end{equation}
where $|P|$ is the pump electric field amplitude. For GaAs parameters
($k_F \sim 1\mu\text{m}^{-1}$, $\omega_0\sim1.5\,\mathrm{eV}/\hbar$, 
$d_0\sim1\,\mu\mathrm{m}$), $\alpha\sim 1$ is achieved at intracavity 
field amplitudes $|P|\sim3.5\times10^7\,\mathrm{V/m}$, 
corresponding to pump powers of order $1\,\mathrm{mW}$ focused to a 
$(10\,\mu\mathrm{m})^2$ spot in a high-finesse cavity 
($\mathcal{F}\sim10^5$), consistent with current photon BEC experiments~\cite{S10,S11,S12}. 

\begin{table}[t!]
\begin{ruledtabular}
\begin{tabular}{cc}
Physical variable & Normalization value \\
\hline
length & $1/k_F$ \\
wavevector & $k_F$ \\
energy & $\varepsilon_0 \equiv \hbar^2 k_F^2/M$ \\
time & $ t_0 \equiv M/ (\hbar k_F^2) $ \\
frequency & $ \hbar k_F^2 / M$ \\
electric field &  $|P|$ \\
$g(\mathbf k)$ & $e^4  /(2\pi \hbar \epsilon_0 d_0  \omega_0^3 ) $ 
\end{tabular}
\end{ruledtabular}
\caption{Normalization values for the relevant physical variables. For a InGaAs microcavity with $M \sim 10^{-5}\,m_e$ and
$k_F \sim 1$--$10\,\mu\mathrm{m}^{-1}$, the energy normalization corresponds to
$\varepsilon_0 \sim 0.1$--$10\,\mathrm{meV}$ and the time to $t_0 \sim  0.1$--$10\,\mathrm{ps}$.}
\label{tab_normalization}
\end{table}

\section{Linear Stability Analysis}\label{LSA}

Using the Bogoliubov decomposition $\mathcal E = (\mathcal E_0 + \delta u_{\mathbf k} \,e^{i \mathbf k \cdot \mathbf r -i\omega t} + \delta v_{\mathbf k}^\ast\,e^{-i\mathbf k \cdot \mathbf r+i\omega t})$ with $N_0 = |\mathcal E_0|^2$ the average photon density, it follows from Eq.~\eqref{GP_normalized} that
\begin{equation}
	\Big(-\Omega_p - i\kappa + \alpha N_0 g_0 \Big) \mathcal E_0+ \Gamma = 0, \label{cubic}
\end{equation}
where we defined $g_0 \equiv g(\mathbf k = 0)$. After taking the modulus we get
\begin{equation}
	\alpha^2 g_0^2 N_0^3 - 2  \Omega_p\alpha  g_0 N_0^2  + (\Omega_p^2 + \kappa^2)N_0 - \Gamma^2 = 0. \label{cubic_N0}
\end{equation}
For fixed ($\alpha$, $\kappa$) and drive parameters ($\Omega_p$, $\Gamma$), Eq.~\eqref{cubic_N0} may have either one or three real solutions, as shown in Fig.~(1) of the main text. \\

\begin{figure}[!t]
\centering
\includegraphics[width=\columnwidth]{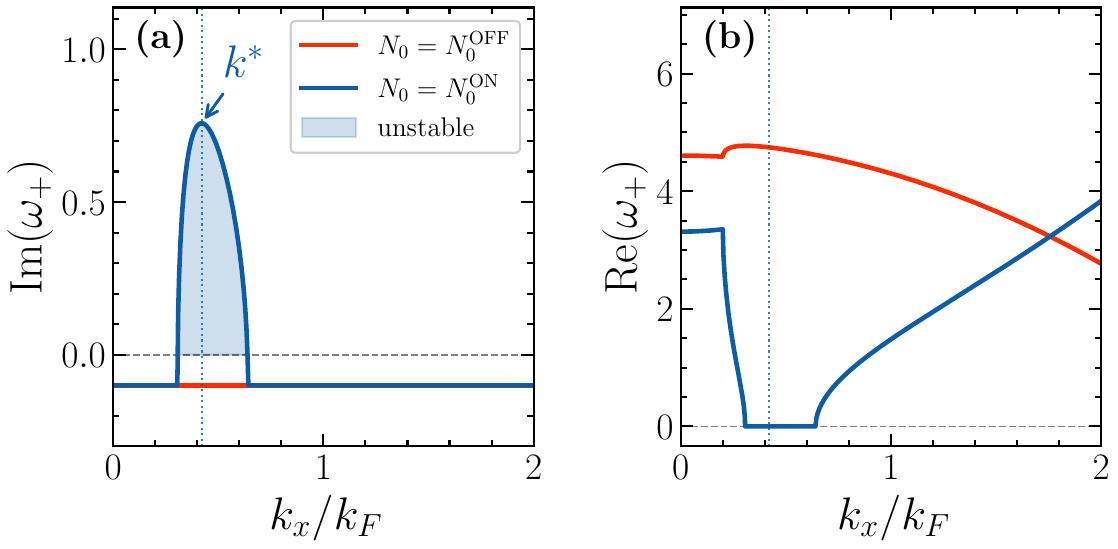}
\caption{Bogoliubov dispersion relation $\omega_+(k_x)$ along the drift
direction for the OFF (lower, $\Nph^\mathrm{OFF}\simeq0.097$) and ON
(upper, $\Nph^\mathrm{ON}\simeq2.935$) branches
($\alpha=1$, $\kappa=0.1$, $\Omega_p=5$, $v_0=0.45\,\vF$).
(a)~Imaginary part $\mathrm{Im}(\omega_+)$: the OFF branch is uniformly
damped at $-\kappa$ (stable), while the ON branch develops a positive-amplitude
band around $\kst\simeq0.42\,\kF$ (shaded), signalling the roton instability.
The peak growth rate $ \Gamma_\mathrm{rot} =0.758\,t_0^{-1}$ is indicated.
(b)~Real part $\mathrm{Re}(\omega_+)$: the ON-branch dispersion shows the
characteristic roton softening near $\kst$, while the OFF branch retains a
conventional phonon-like dispersion.}
\label{fig:SM_bog}
\end{figure}

Linearizing in $\delta u$ and $\delta v$ yields the Bogoliubov eigenvalue problem
\begin{equation}
  \omega
  \begin{pmatrix}\delta u_\mathbf{k} \\ \delta v_\mathbf{k}^*\end{pmatrix}
  = \begin{pmatrix}
    \eta(\mathbf{k})  - i\kappa & \alpha N_0 g(\mathbf k) \\
   -\alpha N_0 g(\mathbf k)  & -\eta(\mathbf{k}) - i\kappa
  \end{pmatrix} 
  \begin{pmatrix}\delta u_\mathbf{k} \\ \delta v_\mathbf{k}^*\end{pmatrix},
  \label{eq:SM_Bog}
\end{equation}
with $\eta(\mathbf{k}) = k^2/2 - \Omega_p + \alpha N_0 g_0$. From Eq.~\eqref{eq:SM_Bog} we obtain the eigenfrequencies
\begin{equation}
  \omega_\pm(\mathbf{k}) = -i\kappa
  \pm\sqrt{\eta(\mathbf{k})\bigl[\eta(\mathbf{k})+2\alpha N_0  g (\mathbf k) \bigr]}.
  \label{eq:SM_dispersion}
\end{equation}
The uniform state is linearly stable if
$\mathrm{Im}(\omega_+) \leq 0$ for all $\mathbf{k}$, i.e.,
\begin{equation}
 \eta(\mathbf{k})\bigl[\eta(\mathbf{k})+2\alpha N_0  g (\mathbf k) \bigr] \geq -\kappa^2
  \quad \forall\,\mathbf{k}.
  \label{eq:SM_stability}
\end{equation}
Violation of the condition~\eqref{eq:SM_stability} at a specific $\mathbf{k}= k^\ast$ signals a roton instability with growth rate 
\begin{equation}
	\Gamma(k^\ast) = \sqrt{|\eta(k^\ast)[\eta(k^\ast)+2\alpha N_0 g(k^\ast)]|-\kappa^2}.
\end{equation}
The roton growth rate for the ON state -- defined in the main text to correspond to $\Nph^\mathrm{ON}\simeq 2.935$ -- is
\begin{equation}
  \Gamma_\mathrm{rot} 
  \simeq 7.6\,\kappa,
  \label{eq:SM_Gbog}
\end{equation}
giving a crystallization time $\tau_e = 1/ \Gamma_\mathrm{rot} 
\simeq 1.3\,t_0$.
Figure~\ref{fig:SM_bog} shows the full Bogoliubov dispersion
$\omega_\pm(k_x)$ for both branches, confirming these analytic results.

\section{Extinction ratio}

The extinction ratio is defined as
\begin{equation}
  \mathrm{ER}(t)
  = 10\log_{10}\!\left(\frac{I_\mathrm{Bragg}(t)}{I_\mathrm{through}(t)}\right),
  \label{eq:SM_ER}
\end{equation}
where the Bragg intensity integrates the spectral power in an annular
window of half-width $\delta k$ around $|\vk|=\kst$:
\begin{equation}
  I_\mathrm{Bragg}(t)
  = \sum_{\bigl||\vk|-\kst\bigr|<\delta k}
    \left|\Ep(\vk,t)\right|^2,
  \label{eq:SM_IBragg}
\end{equation}
and $I_\mathrm{through}$ is the power in a disk of radius
$\delta k$ around $\vk= 0 $:
\begin{equation}
  I_\mathrm{through}(t)
  = \sum_{|\vk|<\delta k}
    \left|\Ep(\vk,t)\right|^2.
  \label{eq:SM_Ithrough}
\end{equation}
Here $\Ep(\vk,t)$ is the 2D spatial Fourier transform of the intracavity field at time $t$. We verified numerically that both $I_\mathrm{Bragg}$ and $I_\mathrm{through}$ are not affected by $\delta k$ provided $\delta k > 0.2$. \par 

The switching contrast 
\begin{equation}
	\Delta\mathrm{ER} \equiv \mathrm{ER}^\mathrm{ON}-\mathrm{ER}^\mathrm{OFF} \quad \text{(in dB)}
\end{equation}
corresponds to a multiplicative gain of $10^{\Delta\mathrm{ER}/10}$ 
in the Bragg-to-forward power ratio upon switching. In the present context, a large 
$\Delta\mathrm{ER}$ is the spectral signature of the superfluid-to-supersolid phase 
transition accompanying the OFF$\,\to\,$ON switch. The supersolid order parameter 
redistributes macroscopic photon population from the forward channel into the Bragg 
modes, while the featureless superfluid leaves those modes dark.

\section{Numerical Method}

All simulations that solve Eq.~\eqref{GP_normalized} use the
Strang split-step Fourier method~\cite{S5,S6} with time step
$dt = 10^{-3}$ on a square periodic domain of side $150\,\kF^{-1}$
discretized on an $256\times256$ grid. The split-step scheme alternates between:
\begin{itemize}
	\item[(i)] a half-step nonlinear advance
$\Ep \to e^{(V_\mathrm{nl}-\kappa)\,dt/2}\Ep - i\Gamma\,dt/2$, where $V_\mathrm{nl} = g*|\Ep|^2$ the nonlinear potential evaluated with a Fast Fourier Transform (FFT) algorithm,
	\item[(ii)]  a full linear propagator $e^{-i(k^2/2-\Omega_p)\,dt}$
applied in Fourier space, 
	\item[(iii)]  a second half-step nonlinear advance identical to (i).
\end{itemize} 

For the initial condition we use profiles of the form $\Ep (\mathbf r, t=0) = \big[\sqrt{N_0} + \xi(\mathbf r) \big] e^{i\psi(\mathbf r)}$ with different $N_0$ depending on the situation. Here $\xi(\mathbf r) \ll \sqrt{N_0}$ is a small random noise with zero mean, i.e., $\langle \xi(\mathbf r)\rangle = 0$, and $\psi(\mathbf r)$ is a random phase uniformly distributed in the interval $[-\pi,\pi[$.

The substitution $\mathcal{E}(\mathbf{r},t)\to\mathcal{E}(\mathbf{r},t)e^{-i\Omega_p t}$ in the simulated protocols provides solutions consistent with Eq.~\eqref{cubic_N0} for any envelope satisfying $|\dot{\mathcal{E}}/\mathcal{E}|\ll\Omega_p$. For ps-scale pulses and an optical pump ($\Omega_p\sim200$~THz), this ratio is $\lesssim10^{-3}$, so the slow-varying envelop approximation holds by orders of magnitude. The pump amplitude may equally be time-dependent, $\Gamma(t)=\Gamma_0(t)e^{-i\Omega_p t}$, provided $|\dot{\Gamma}_0/\Gamma_0|\ll\Omega_p$---satisfied for any modulation on ps--ns timescales. Moreover, the step-function write/erase pulses used in the simulations are valid for any physical pulse with finite rise time $\tau_\text{rise}\ll\kappa^{-1}$ provided $\Gamma_\text{wr}>\Gamma_\text{max}$ during the pulse and the pulse duration exceeds $t^*_\text{wr}$ (see Sec.~\ref{sec_WPDT}).

\section{Experimental Parameters for $\text{GaAs}$ Microcavities}

Table~\ref{tab:SM_params} lists representative physical parameters
for a GaAs--based implementation of the proposed switch.

\begin{table}[h]
\begin{ruledtabular}
\begin{tabular}{lccc}
Parameter & Symbol & Value & Sim.\ unit \\
\hline
Fermi wavevector   & $\kF$            & 1--10~$\mu$m$^{-1}$           & (def.)      \\
Fermi velocity     & $\vF$            & $10^4$--$10^5$~m/s            & (def.)      \\
Photon mass        & $m$              & $\sim10^{-5}\,m_e$            & ---         \\
Cavity Q-factor    & $Q$              & $10^4$--$10^5$                & ---         \\
Photon lifetime    & $\tau_c$         & 1--100~ps              & $\kappa=0.1$\\
Pump detuning      & $\Omega_p$       & $\sim$1~meV                   & $\Omega_p=5$\\
Electron density   & $n_e$            & $10^{10}$--$10^{11}$~cm$^{-2}$& ---        \\
Drift velocity     & $v_0$            & $0.45\,\vF$                   & $0.45$  \\
Roton wavevector   & $\kst$           & $0.42\,\kF$                   & $0.42$      \\
Lattice period     & $a=2\pi/\kst$    & 1.5--15~$\mu$m         & $\simeq15$ \\
Switching time     & $\tau_\mathrm{sw}$& 5--500~ps            & $\sim50$    \\
\end{tabular}
\end{ruledtabular}
\caption{Representative physical parameters for a GaAs DBR microcavity
with an embedded 2DEG.
When the corresponding dimensionless parameters is required or determines some required input value for the simulations, the associated simulation parameter is given in the rightmost column.
The third column spans the full range $\kF = 1$--$10\,\mu$m$^{-1}$, while the physical numbers quoted in the main text all refer to the single reference design $\kF \simeq 3\,\mu$m$^{-1}$, for which $t_0 \simeq 1$~ps, $\tau_c = 10\,t_0 \simeq 10$~ps, $\tau_\mathrm{sw} = 50\,t_0 \simeq 50$~ps and $a \simeq 5\,\mu$m.}
\label{tab:SM_params}
\end{table}

The cavity Q-factor sets the photon lifetime $\tau_c=Q/\omega_c$.
For $\omega_0\simeq1.5$~eV/$\hbar$ and
$Q=10^4$--$10^5$, $\tau_c\simeq4$--$40$~ps, corresponding to
$\kappa\simeq0.025$--$0.25$~ps$^{-1}$.
The electron density $n_e$ sets $\kF=\sqrt{2\pi n_e}$. For the reference
design $\kF\simeq3\,\mu$m$^{-1}$ this gives a lattice period
$a=2\pi/\kst\simeq5\,\mu$m and a time unit $t_0\simeq1$~ps, so that the
switching time is $\tau_\mathrm{sw}=50\,t_0\simeq50$~ps.

The drift velocity $v_0=0.45\,\vF$ is achieved by applying a
voltage of order $\Delta V = m^*v_0^2/(2e) \sim 1$~mV (with
$m^*\simeq0.067\,m_e$ the GaAs electron effective mass) across
the 2DEG QW, within reach of standard gate-voltage control~\cite{S7,S8}.
Rotating the drift direction by $90^\circ$ requires only reversing
the gate-bias polarity between two pairs of orthogonal contacts,
with no optical realignment.

In the cubic Eq.~(4) the nonlinear coefficient $\alpha g_0$ involves the zero-momentum Lindhard polarizability, which is derived assuming a sharp Fermi surface, i.e.\ a mean free path $\ell = v_F\tau_e \gg k_F^{-1}$. For GaAs 2DEGs with $\mu_e \gtrsim 10^5$\,cm$^2$/V$\cdot$s, $\ell\sim10$--$100\,\mu$m against $k_F^{-1}\sim1\,\mu$m, so this condition is comfortably met. More critically, the attraction at $k^*$ that drives the roton instability of Eq.~(5) originates from the Kohn anomaly in the real part of the Lindhard susceptibility, which sharpens with increasing mobility. Higher mobility therefore deepens the roton dip, strengthens the ON-branch supersolid order, and enhances the switching contrast. State-of-the-art GaAs 2DEGs reach $\mu_e\sim10^6$\,cm$^2$/V$\cdot$s at low temperature~\cite{S7,S8}, placing this platform well within the high-mobility regime required.

The switch state is read by collecting the far-field emission at
polar angle $\theta_B = \arcsin(\hbar\kst/p_\mathrm{ph})$ from the
cavity axis, where $p_\mathrm{ph} \simeq \hbar\omega_0/c$
is the in-plane photon momentum at the cavity mode angle.
For $\kst\simeq1\,\mu$m$^{-1}$ and $\omega_0\simeq1.5$~eV/$\hbar$
one obtains $\theta_B\simeq0.3^\circ$, easily resolved by standard
confocal microscopy.

\subsection{Photon--photon coupling estimation}

\begin{figure}[t!]
\centering
\includegraphics[width=\columnwidth]{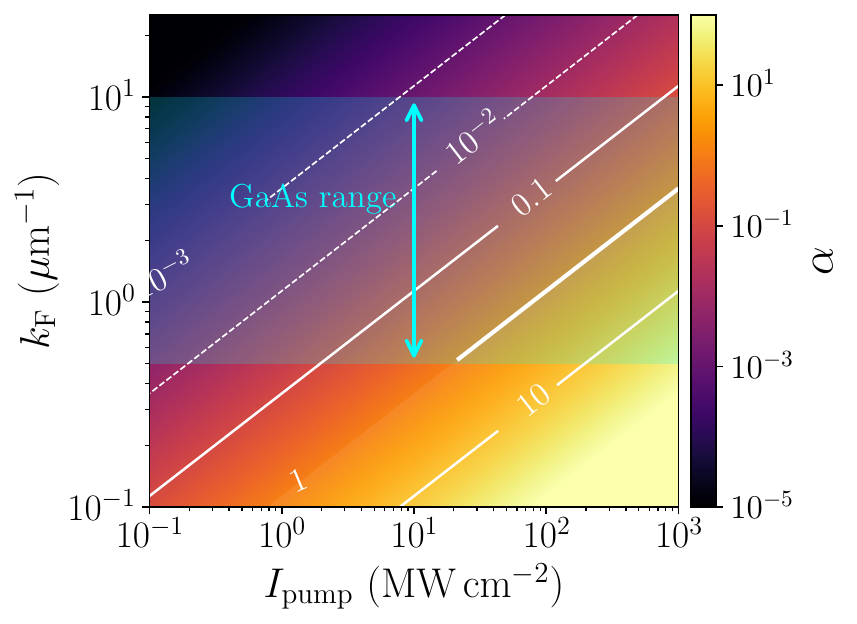}
\caption{Photon--photon coupling as a function of Fermi wavevector and pumping intensity. Fixed parameters $d_0 = 0.5 \mu$m and $\omega_0 = 1.5$ eV$/\hbar$. }
\label{fig_alpha}
\end{figure}

Considering $\omega_0 \approx 1.5$\,eV/$\hbar$,
$\kF = 1\,\mu\mathrm{m}^{-1}$, and $d_0 = 0.5 \,\mu$m,
the threshold pump field corresponding to $\alpha = 1$ is
\begin{equation}
  |P| = \sqrt{\frac{2\pi\hbar^3\kF^2\epsilon_0 d_0\omega_0^3}{e^4}}
        \simeq 2.5\times10^7\,\mathrm{V\,m}^{-1},
  \label{eq:SM_P0}
\end{equation}
or equivalently, the threshold pump intensity is
$I_0 = \epsilon_0 c\,|P_0|^2/2 \simeq 78$\,MW\,cm$^{-2}$.
This intensity is accessible in the pulsed regime
(e.g.\ 100\,ps pulses at a fluence $\sim 1.5$\,mJ\,cm$^{-2}$,
well below the GaAs damage threshold of $\sim 100$\,mJ\,cm$^{-2}$
for picosecond pulses~\cite{S7}).
At the more moderate pulsed intensity of $1$\,MW\,cm$^{-2}$
one obtains $\alpha \approx 0.064$, which still lies in the
bistable regime provided the pump detuning satisfies $\Omega_p > \sqrt{3}\,\kappa$. \\
Equation~\eqref{eq:alpha} reveals three practical handles on $\alpha$:
\begin{itemize}
  \item \textit{2DEG density}: $\alpha \propto \kF^{-2} \propto n_e^{-1}$, so
        sparser electron sheets (lower gate voltage) enhance the coupling.
        For $\kF = 0.5\,\mu\mathrm{m}^{-1}$ at $1$\,MW\,cm$^{-2}$,
        $\alpha \approx 0.26$.
  \item \textit{Cavity thickness}: $\alpha \propto d_0^{-1}$; a thinner
        spacer (e.g.\ $d_0 = 50$\,nm) doubles $\alpha$ relative to 100\,nm.
  \item \textit{Pump power}: $\alpha \propto |P|^2 \propto I_\mathrm{pump}$,
        so $\alpha$ is tunable continuously from zero up to the
        damage-limited ceiling.
\end{itemize}
Figure~\ref{fig_alpha} shows the values of $\alpha$ for our range of interest. 
Combining $\kF = 0.5\,\mu\mathrm{m}^{-1}$ and $d_0 = 50$\,nm,
$\alpha \approx 1$ is achievable at pump intensities of order
$1$\,MW\,cm$^{-2}$, well within the reach of standard pulsed
semiconductor lasers.
The simulations in the main text use $\alpha = 1$ as a representative
value in this accessible regime.

\section{Roton-Mode Exponential Growth}

To directly validate the roton growth rate $ \Gamma_\mathrm{rot} $ given in Sec.~\ref{LSA}, we initialize the simulation at the upper-branch homogeneous state $\mathcal E_0$ such that $|\mathcal E_0|^2 = \Nph^\mathrm{ON}\simeq2.935$
as a result of the cubic equation at $\Gamma=1.5$, add
isotropic Gaussian noise of amplitude $\Delta \mathcal E =10^{-2}|\Ep_0|$,
and evolve under the hold drive $\Gamma_\text{hold}=1.5$ until saturation is reached. The structure factor at the roton wavevector,
\begin{equation}
  S(\kst,t) = \frac{1}{N^4}
  \sum_{\bigl||\vk|-\kst\bigr|<\delta k}
  \bigl|n(\vk,t)\bigr|^2,
  \label{eq:SM_Skstar}
\end{equation}
with $n(\vk,t)$ the Fourier transform of
$|\Ep(\bvr,t)|^2 - |\mathcal E_0|^2$ and
$\delta k = 0.1$, is recorded at every time step.

\begin{figure}[t!]
\centering
\includegraphics[width=\columnwidth]{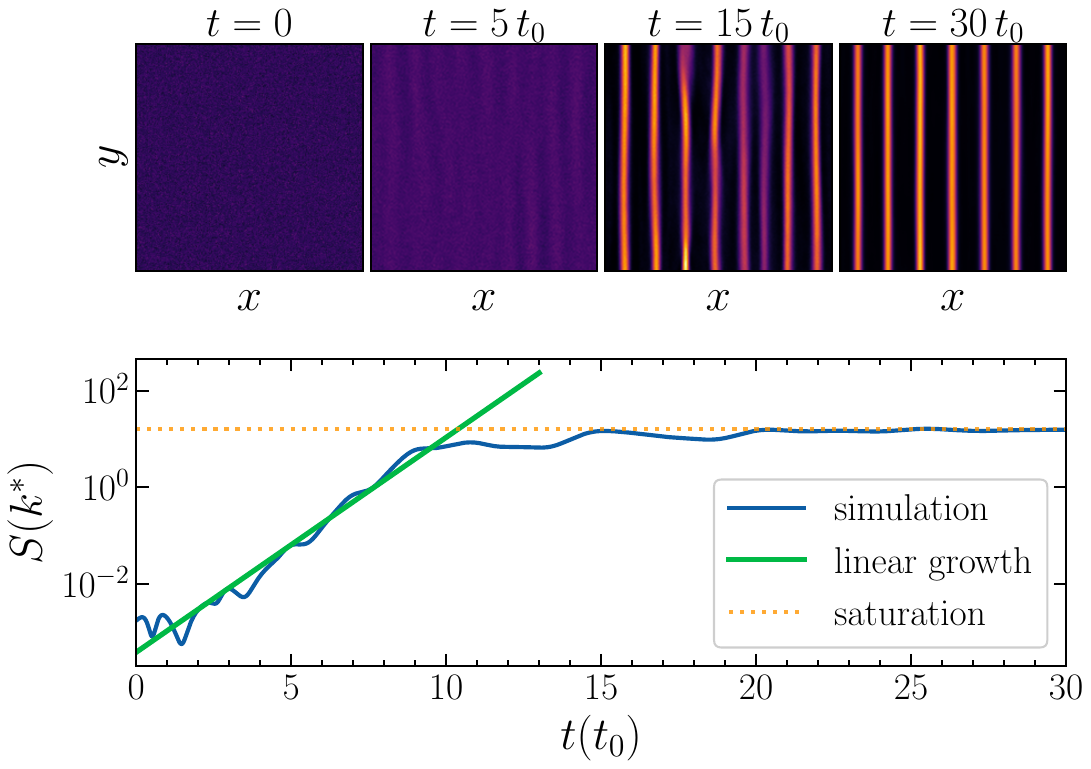}
\caption{Roton-mode exponential growth.
Top row: real-space photon density $|\Ep(\bvr)|^2$
at $t = (0,\,5,\,15,\,3
0)\,t_0$, showing the build-up of the stripe
supersolid from an initially homogeneous state.
Bottom: $S(\kst,t)$ (solid blue) vs.\ time.
The dashed green line is the Bogoliubov prediction
$\sim e^{2 \Gamma_\mathrm{rot} t}$ with $ \Gamma_\mathrm{rot} =0.758\,t_0^{-1}$;
the dotted orange line marks the supersolid saturation value
$S^\mathrm{ss}\simeq 
16$.
Parameters: $\alpha=1$, $\kappa=0.1$, $\Omega_p=5$, $\Gamma=1.5$.}
\label{fig:SM_growth}
\end{figure}

Figure~\ref{fig:SM_growth} shows $S(\kst,t)$ on a logarithmic scale.
Three distinct regimes are visible:
(i)~an initial \emph{linear-amplification} window ($t\lesssim 10\,t_0$)
where $S(\kst,t)\propto\exp(2 \Gamma_\mathrm{rot} \,t)$,
consistent with the Bogoliubov prediction
(for comparison, we show the theoretical line with slope $2 \Gamma_\mathrm{rot} =1.516$ in dimensionless units);
(ii)~a \emph{nonlinear-saturation} crossover
($10\lesssim t\lesssim 20\,t_0$) in which competing Bragg modes
interact and the growth slows;
(iii)~a \emph{supersolid steady state} ($t\gtrsim 20\,t_0$)
at $S(\kst)\simeq 16$, which corresponds to dynamic range of
$\sim 10^2$~dB above the initial noise floor $ S(\kst,0)$.

\section{Write-Pulse Duration Threshold}\label{sec_WPDT}
A key practical figure of merit is the minimum write-pulse duration
$t_\mathrm{wr}^*$, defined as the shortest pulse that reliably places the system
to the supersolid ON state. Next, we quantify this by scanning $t_\mathrm{wr}$ at fixed
$\Gamma_\mathrm{wr}=3.5$ and measuring the structure factor
$S(\kst)$ after saturation due to the hold drive $\Gamma_\text{hold}=1.5$.\\

\begin{figure}[t!]
\centering
\includegraphics[scale=0.6]{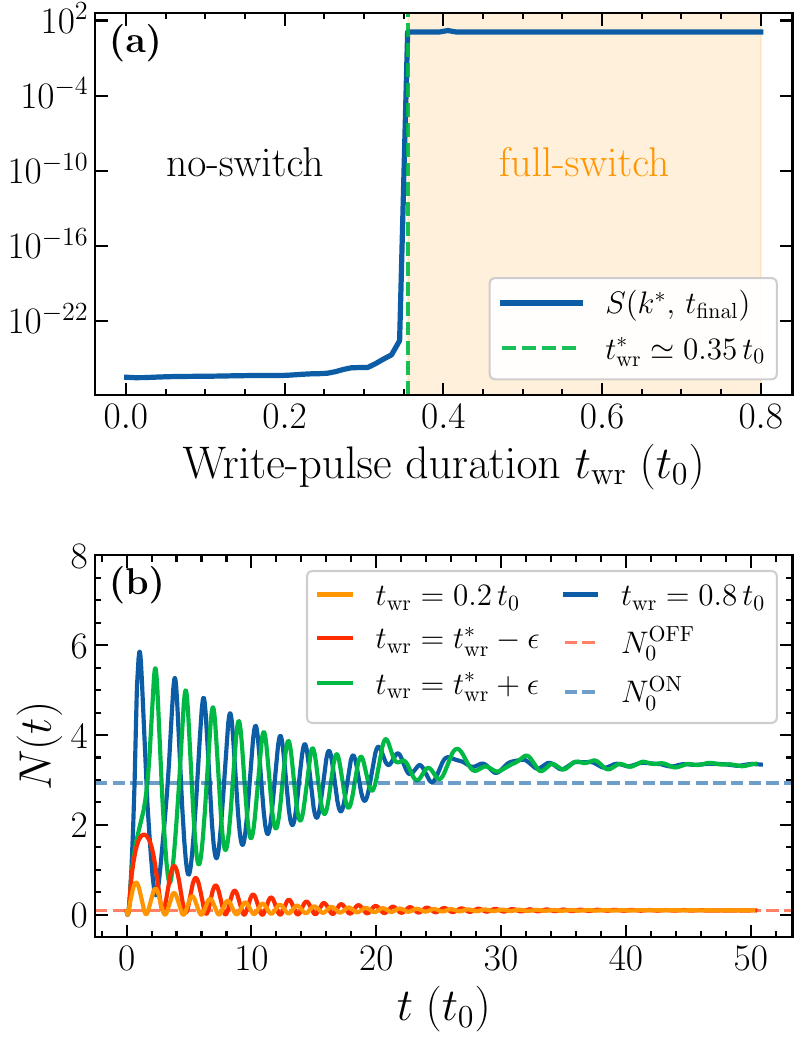}
\caption{Write-pulse duration characterization.
(a)~Structure factor $S(\kst)$ measured after $200\,t_0$ of hold drive
as a function of write-pulse duration $t_\mathrm{wr}$
($\Gamma_\mathrm{wr}=3.5$, $\Gamma_\mathrm{hold}=1.5$).
A sharp switching threshold at $t^*_\mathrm{wr} \simeq 0.35\,t_0$ 
(dotted green line) separates no-switch ($S \simeq 0$) from full-switch 
($S \simeq 12.2$) outcomes. (b)~Mean photon density $N(t)$ for four representative pulse durations: below threshold ($t_\mathrm{wr} = 0.2 t_0$), at threshold from below ($t_\mathrm{wr} = t^*_\mathrm{wr} - \epsilon$), at threshold from above ($t_\mathrm{wr} = t^*_\mathrm{wr} + \epsilon$) and above threshold ($t_\mathrm{wr} = 0.8 t_0$), with $\epsilon$ a small positive number.}
\label{fig:SM_twrite}
\end{figure}

The results are shown in Fig.~\ref{fig:SM_twrite}.
As expected, for a pulse duration $t_\mathrm{wr} < t_\mathrm{wr}^*$ the field does not reach the
upper branch before the pulse ends, and the system relaxes back to the OFF
superfluid. For $t_\mathrm{wr} \geq t_\mathrm{wr}^*$, the system commits to the
upper branch and the supersolid nucleates at $S(\kst)\simeq12.2$ regardless of further pulse lengthening. We find $t^*_\mathrm{wr} \simeq 0.38\,t_0 \simeq 0.04\,\kappa^{-1}$, i.e.\
the write pulse need only last a small fraction of the photon lifetime.\\

The physical reason behind the existence of a nonzero value of $t_\mathrm{wr}^*$ is that the photon field is driven above the upper saddle-node during the write pulse, causing $N(t)$ to increase on the loss timescale $\kappa^{-1}$. Hence, a pulse duration $t_\mathrm{wr}\gtrsim
\kappa^{-1}/5$ is required to push $N(t)$ past the upper limit $\Gamma_\text{max}$ so that
after the pulse the hold drive sustains the upper branch. The subsequent supersolid formation time ($\sim 20\,t_0$) then proceeds autonomously, governed by the Bogoliubov growth. 
The minimum (dimensionless) write-pulse energy is
\begin{equation}
  \Delta E_\mathrm{min}  = |\Gamma_\mathrm{wr}|^2\,t_\mathrm{wr}^*
\simeq 5.1 , 
  \label{eq:SM_Emin}
\end{equation}
which is consistent with the
sub-fJ switching energy estimated in the main text for GaAs since $t_\mathrm{wr}^*\sim0.4\,\mathrm{ps}$,

\begin{figure}[t!]
\centering
\includegraphics[width=\columnwidth]{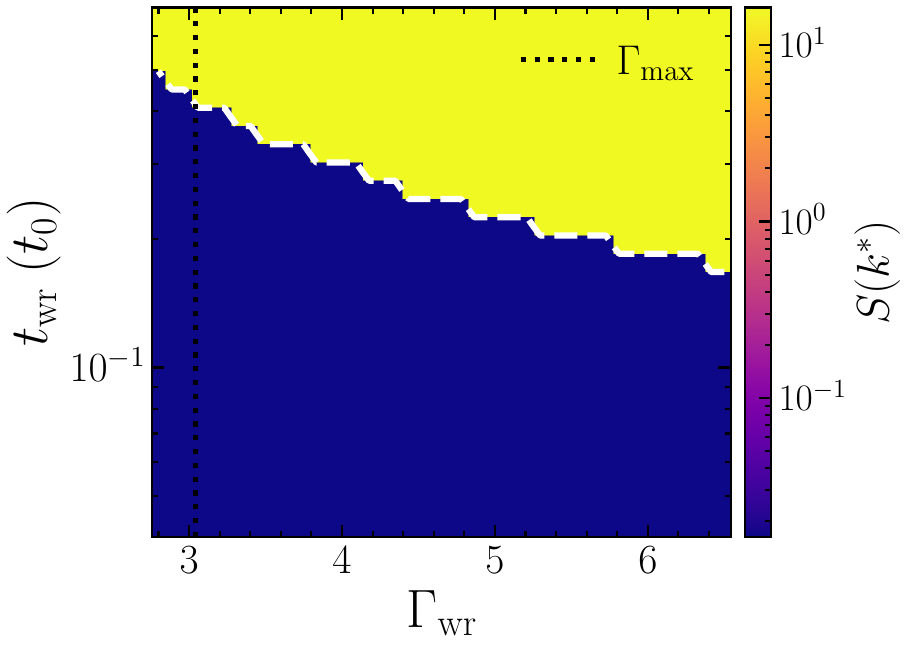}
\caption{Write-pulse phase diagram. Colour map of the final structure factor $S(\kst)$ in the $(\Gamma_\mathrm{wr},\,t_\mathrm{wr})$ plane ($t_\mathrm{hold}=100\,t_0$). Yellow: switched (supersolid ON state); blue: no switch. White dashed contour: switching threshold $t^*_\mathrm{wr}(\Gamma_\mathrm{wr})$. Black dotted line: upper bistable fold $\Gamma_\mathrm{max}\approx3.04$. The monotonically decreasing boundary confirms that larger $\Gamma_\mathrm{wr}$ reduces the required pulse duration.}
\label{fig:SM_phasediag}
\end{figure}

Figure~\ref{fig:SM_phasediag} generalises this result to the full $(\Gamma_\mathrm{wr},\,t_\mathrm{wr})$ plane. The switching boundary decreases monotonically with write-pulse amplitude and remains well below one photon lifetime throughout the range explored, confirming that the fixed-$\Gamma_\mathrm{wr}$ scan above captures the generic behaviour. The switching is possible for $\Gamma_\mathrm{wr}>\Gamma_\mathrm{max}$ regardless of pulse length, consistent with the bistable saddle-node threshold.

\section{Robustness Under Parameter Variations}
To assess the practical tolerance of the switching scheme, we vary both the write-pulse amplitude
$\Gamma_\mathrm{wr}$ and the hold-drive amplitude
$\Gamma_\mathrm{hold}$ while keeping all other parameters fixed.

\subsection{Write-amplitude threshold}
Figure~\ref{fig:SM_params}(a) shows $N(t)$ for different values of 
$\Gamma_\mathrm{wr}$
and fixed $\Gamma_\mathrm{hold}=1.5$, showing that 
drives at $\Gamma_\mathrm{wr} \in 2 - 2.5$ fail to switch the system:
after the pulse, $N$ returns to the OFF branch [$N\simeq0.097$,
$S(\kst)\simeq 0$].
For $\Gamma_\mathrm{wr}\geq 3.0$ the system reliably switches to the ON
state and the supersolid forms.
The effective threshold therefore lies between $\Gamma=2.5$ and $3.0$,
close to the upper bistable limit as expected.
This confirms that the switching mechanism is the bistable saddle-node:
for drives below $\Gamma_\text{min}$, the field does not reach the upper branch within the
$15\,t_0$ write window; for drives at or above $\Gamma_\text{max}$, the lower branch ceases
to exist and the field is forced to the ON state unconditionally.
Crucially, the final ON-state density $N^\mathrm{ON}(t\rightarrow \infty) \simeq 3.34$
is independent of $\Gamma_\mathrm{wr}$ for all switched cases,
confirming that it is determined by the hold-drive steady state, not by the
write amplitude.

\subsection{Hold-drive fidelity}
\begin{figure*}[t]
\centering
\includegraphics[scale=0.65]{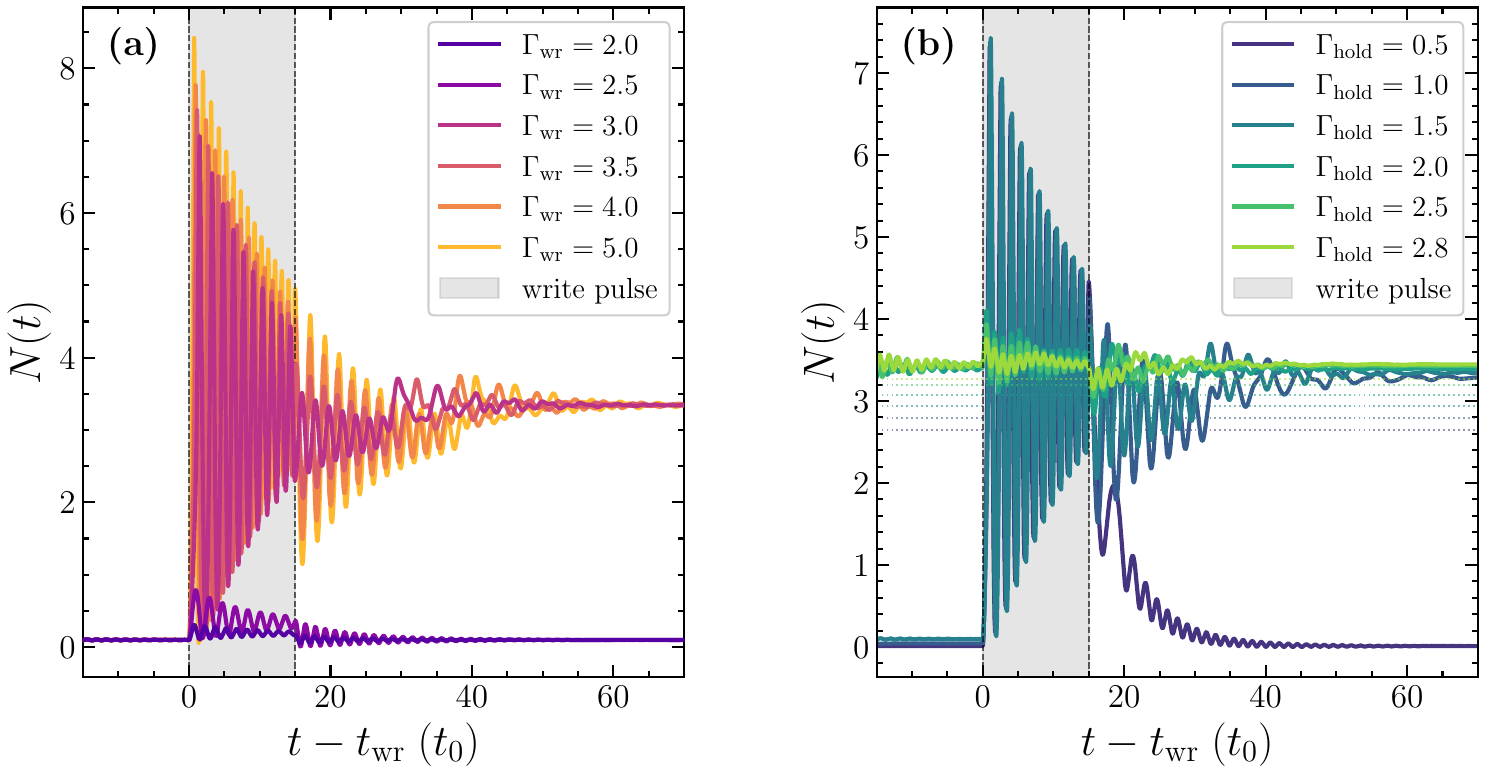}
\caption{Parameter robustness of the write--hold switching scheme
(a)~Mean photon density $N(t)$ for varying write amplitude
$\Gamma_\mathrm{wr}$ at fixed $\Gamma_\mathrm{hold}=1.5$.
(b)~$N(t)$ for varying hold amplitude $\Gamma_\mathrm{hold}$
at fixed $\Gamma_\mathrm{wr}=3.5$.
All values lie within the bistable window~$[0.16,\,3.04]$.
Dotted horizontal lines mark the corresponding cubic upper-branch densities
$\Nph^\mathrm{ON}(\Gamma_\mathrm{hold})$.
The ON state is retained for $\Gamma_\mathrm{hold}\geq 1.0$, and it fails
at $\Gamma_\mathrm{hold}=0.5$ due to trajectory overshoot across the
unstable middle root.
In both panels the grey shaded band marks the write-pulse interval.}
\label{fig:SM_params}
\end{figure*}

Figure~\ref{fig:SM_params}(b) shows $N(t)$ for varying 
$\Gamma_\mathrm{hold} $
at fixed $\Gamma_\mathrm{wr}=3.5$.
All six values lie inside the bistable window $\Gamma\in[0.16,\,3.04]$.
For $\Gamma_\mathrm{hold}\geq 1.0$ the ON state persists after the write
pulse and the supersolid is maintained: $N(t\gg t_\mathrm{wr})$
settles near the corresponding cubic upper-branch value
(dotted lines), with $S(\kst)$ remaining at 10--14.
For $\Gamma_\mathrm{hold}=0.5$ the memory is not retained: $N(t)$
decays to the OFF branch after the write pulse.
The reason is dynamical rather than static--the upper branch at
$\Gamma=0.5$ is a valid steady-state solution ($N^\mathrm{ON}=2.64$),
but the large transient overshoot ($N\simeq7$) generated by the
write pulse drives the field so far above the upper branch that the
subsequent relaxation trajectory crosses the unstable middle root of the
cubic, attracting the system into the lower-branch.
For $\Gamma_\mathrm{hold}\geq 1.0$, the upper-branch basin is
sufficiently wide to absorb the overshoot.

In summary, the write--hold protocol is robust over a range
of amplitude $\sim 5\times \Gamma_\text{hold}$ ($1.0\lesssim \Gamma_\mathrm{hold}\lesssim 2.8$,
i.e.\ factor ${\simeq}6$--19 above the lower limit $\Gamma_\text{min}$) and requires a write
amplitude within $\simeq 1$~dB of the upper bistable limit.
This tolerance is consistent with the wide bistable $S$-curve discussed in
Sec.~II and accessible in semiconductor microcavity experiments.

\section{Multi-State Supersolid Memory Switch}

The binary write--erase protocol of the main text generalises naturally 
to a three-state switch by embedding a second quantum-well layer whose 
drift direction $\phi_2$ is independently controlled. The OFF state 
remains the uniform superfluid, while two ON states differ solely in 
supersolid symmetry, selected by $\phi_2$ prior to the write pulse 
with no modification to the pump protocol. Any sequence of states in 
$\{\mathrm{OFF},\,\mathrm{ON}_1,\,\mathrm{ON}_2\}$ is accessible, 
including direct $\mathrm{ON}_1\leftrightarrow\mathrm{ON}_2$ 
transitions that bypass the OFF state.

\begin{figure}[!t]
\centering
\includegraphics[scale = 0.35]{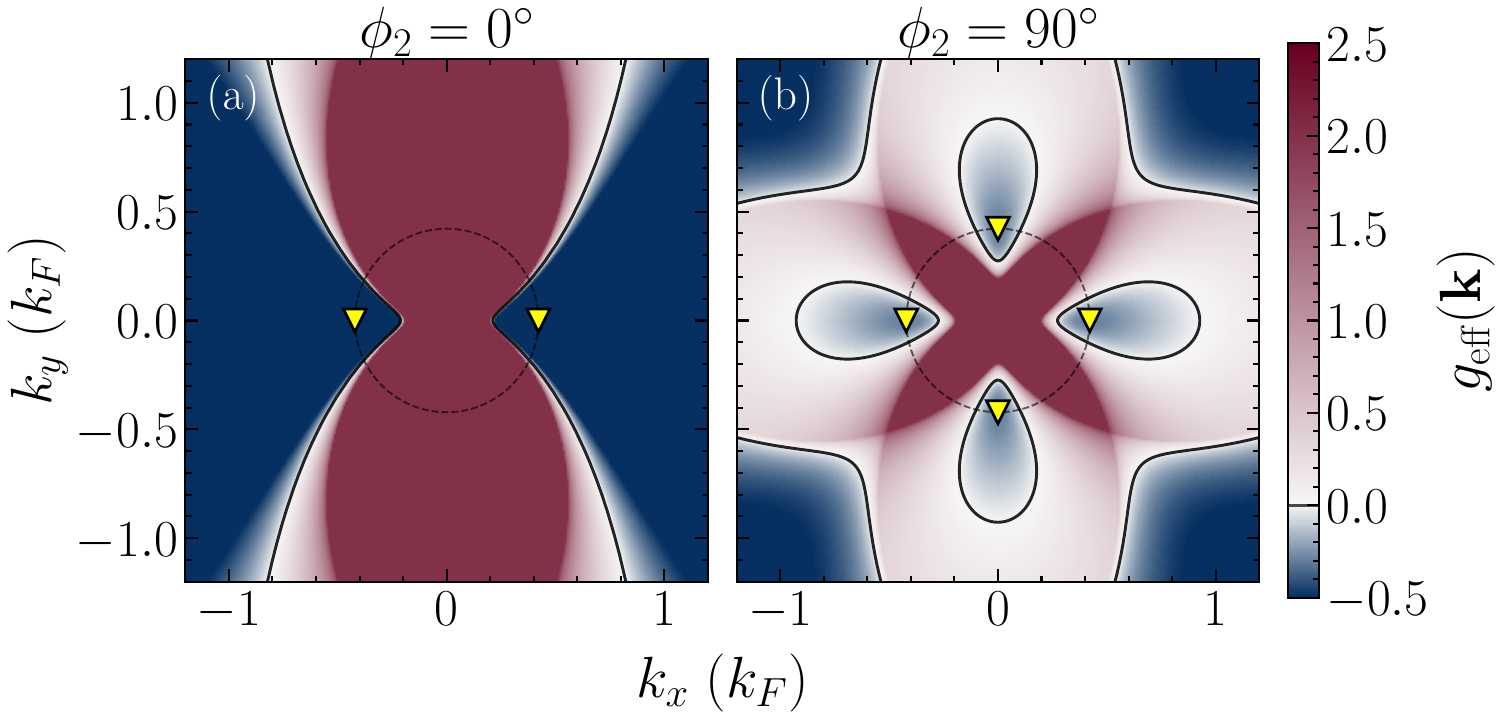}
\caption{Effective interaction kernel $g_\mathrm{eff}(\mathbf{k})$
[Eq.~\eqref{eq:SM_geff_NL}] for the two-layer device
($v_0=0.45\,\vF$).
(a)~Aligned layers ($\phi_2=0^\circ$, ON$_1$): the kernel has a single
pair of deep roton minima at $(\pm\kst,0)$; the $y$-direction is entirely
in the positive (stable) region.
(b)~Perpendicular layers ($\phi_2=90^\circ$, ON$_2$): rotating the
second layer by $90^\circ$ introduces two additional minima at
$(0,\pm\kst)$ (annotated), producing fourfold square symmetry.
Yellow markers indicate roton minima on the ring
$|\mathbf{k}|=\kst$ (dashed circle).}
\label{fig:SM_geff}
\end{figure}

\subsection{Symmetry selection via the effective kernel}

With layer~1 fixed at $\phi_1=0^\circ$ and layer~2 at angle $\phi_2$, 
the intracavity field couples to the combined Lindhard kernel
\begin{equation}
  g_\mathrm{eff}(\mathbf{k};\phi_2)
  = g(\mathbf{k};\phi_1) + g(\mathbf{k};\phi_2) - g_0,
  \label{eq:SM_geff}
\end{equation}
where the subtraction of $g_0$ enforces 
$g_\mathrm{eff}(\mathbf{k}\to 0)=g_0$, leaving the mean-field 
bistability and all switching thresholds unchanged.

\textit{Aligned layers} ($\phi_2=0^\circ$, $\mathrm{ON}_1$).
Both layers reinforce the roton along $k_x$:
$g_\mathrm{eff}(k^*\hat{x};0^\circ)=2g(k^*)-g_0\approx-2.59$,
while $g_\mathrm{eff}(k^*\hat{y};0^\circ)=g_0=+2.0>0$.
The Bogoliubov instability is exclusively along $k_x$, 
nucleating a stripe supersolid with Bragg peaks at $(\pm\kst,0)$.

\textit{Perpendicular layers} ($\phi_2=90^\circ$, $\mathrm{ON}_2$).
Layer~1 drives a roton along $x$ and layer~2 along $y$, yielding
$g_\mathrm{eff}(k^*\hat{x};90^\circ)
=g_\mathrm{eff}(k^*\hat{y};90^\circ)\approx-0.294$.
Both roton modes are degenerate and simultaneously unstable, so the 
write pulse nucleates a square supersolid with four Bragg peaks at 
$(\pm\kst,0)$ and $(0,\pm\kst)$.

Figure~\ref{fig:SM_geff} shows the 2D map of $g_\mathrm{eff}(\mathbf{k})$ 
for both configurations, with the zero contour ($g_\mathrm{eff}=0$) 
highlighted to delineate the unstable regions and the roton minima 
on the ring $|\mathbf{k}|=\kst$ marked explicitly.

\subsection{Three-state write--hold--erase cycle}

Figure~\ref{fig:SM_protocols} (left column) shows a single continuous 
simulation executing three successive cycles under the same pump 
protocol ($\Gamma_\mathrm{wr}=3.5$, $\Gamma_\mathrm{hold}=1.5$, 
$\Gamma_\mathrm{er}=0.05$):
\begin{enumerate}
  \item[(i)]   Cycle~A ($\phi_2=0^\circ$):  
               OFF $\to$ ON$_1$ $\to$ OFF.
  \item[(ii)]  Cycle~B ($\phi_2=90^\circ$, gate bias reversed):  
               OFF $\to$ ON$_2$ $\to$ OFF.
  \item[(iii)] Cycle~C ($\phi_2=0^\circ$ restored):  
               OFF $\to$ ON$_1$, confirming cyclic operation.
\end{enumerate}
The two ON states are distinguished by the structure-factor order 
parameters $S_x\equiv S(\kst,0)$ and $S_y\equiv S(0,\kst)$ 
[Eq.~\eqref{eq:SM_Skstar}], and distinguished from the OFF state by the steady-state value of $N(t)$. The additional diagnostic is the anisotropy ratio:
\begin{align*}
  \mathrm{OFF}:&\quad S_x\approx S_y\approx 0,\\
  \mathrm{ON}_1:&\quad S_x\gg S_y \quad (S_y/S_x\sim10^{-2}),\\
  \mathrm{ON}_2:&\quad S_x\approx S_y \quad (S_y/S_x\sim 1).
\end{align*}
In Cycle~A, $S_x$ saturates after the write pulse while $S_y$ remains 
at the noise floor, consistent with the absence of a $y$-roton for 
$\phi_2=0^\circ$. In Cycle~B, both modes grow simultaneously and 
saturate at comparable values, confirming square-lattice order. Each 
component of $\mathrm{ON}_2$ is approximately 15 times smaller than 
$S_x$ in $\mathrm{ON}_1$, consistent with the order parameter being 
distributed over twice as many Bragg peaks. Cycle~C reproduces the 
$\mathrm{ON}_1$ fingerprint quantitatively, and both order parameters 
return to the noise floor after each erase pulse.

\subsection{Direct state reconfiguration without erasure}

\begin{figure*}[!t]
\centering
\includegraphics[width=\textwidth]{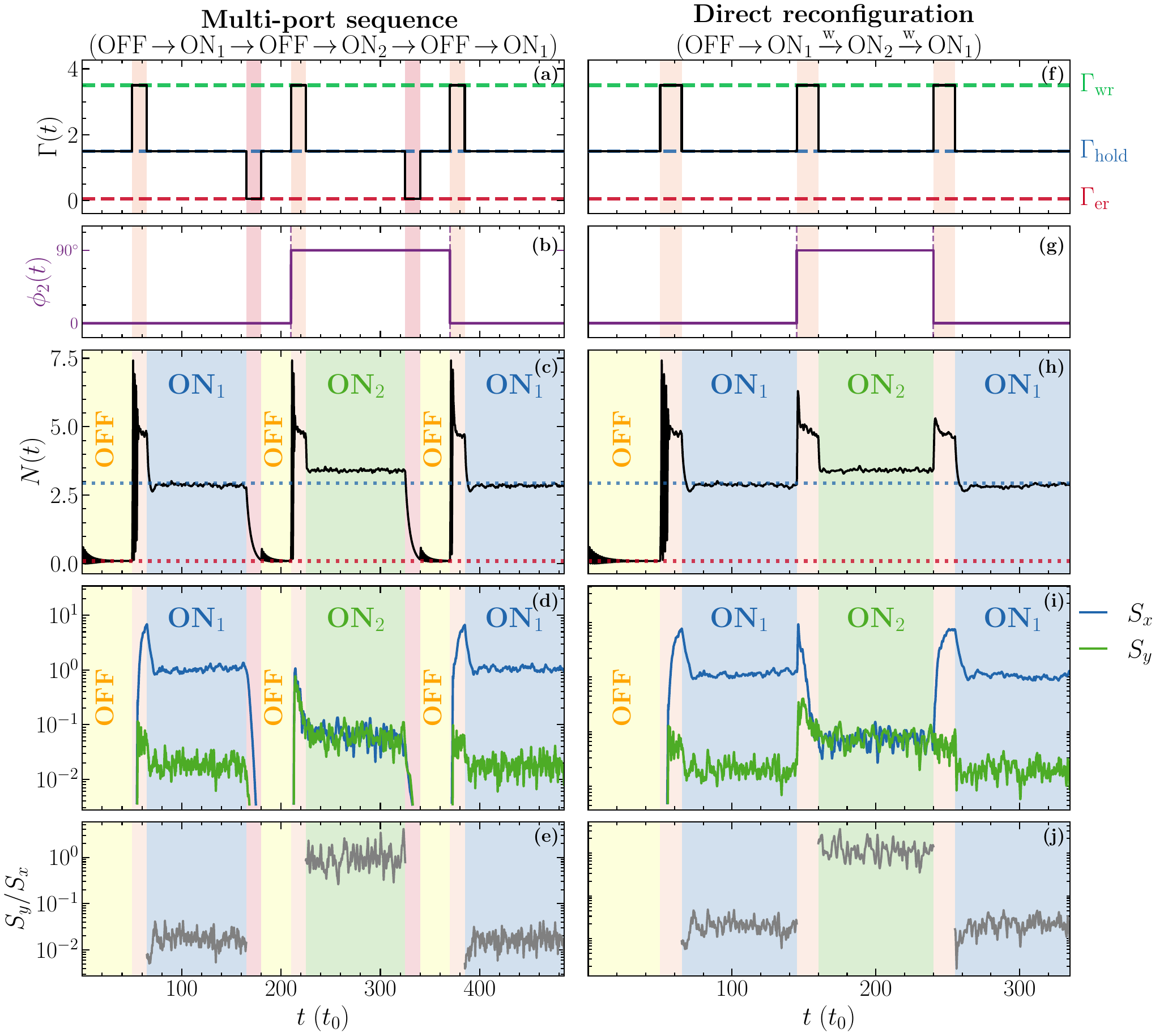}
\caption{Two distinct multi-state switching protocols for the two-layer device.
Blue (green) background shading marks $\mathrm{ON}_1$ ($\mathrm{ON}_2$) hold windows;
orange (red) bands mark write (erase) pulses.
Left column~(a)--(d): write--hold--erase sequence,
$\mathrm{OFF} \to \mathrm{ON}_1 \to \mathrm{OFF} \to \mathrm{ON}_2 \to 
\mathrm{OFF} \to \mathrm{ON}_1$.
Right column~(e)--(h): direct reconfiguration without erase,
$\mathrm{OFF} \to \mathrm{ON}_1 \to \mathrm{ON}_2 \to \mathrm{ON}_1$ 
using only write pulses.
(a,\,e)~Pump amplitude $\Gamma(t)$.
Left: alternating write ($\Gamma_\mathrm{wr}=3.5$, orange) and erase
($\Gamma_\mathrm{er}=0.05$, red) pulses bracketing each hold window;
dotted line: $\Gamma_\mathrm{hold}=1.5$.
Right: three write pulses only.
(b,\,f)~Layer angle $\phi_2(t)$, switched between $0^\circ$ ($\mathrm{ON}_1$) 
and $90^\circ$ ($\mathrm{ON}_2$) by reversing the gate bias.
(c,\,g)~Mean photon density $N(t)$.
Each write pulse drives a transient spike above the ON fixed point
(upper dotted line), while erase pulses return the density to the OFF level 
(lower dotted line). In the right column $N(t)$ remains on the upper branch 
throughout, confirming that the system is never erased.
(d,\,h)~Structure-factor order parameters
$S_x\!\equiv\!S(k^*,0)$ (solid blue) and $S_y\!\equiv\!S(0,k^*)$
(dashed green). In $\mathrm{ON}_1$ ($\phi_2=0^\circ$), $S_x$ saturates while 
$S_y$ remains at the noise floor, giving anisotropy ratio $S_y/S_x \sim 10^{-2}$. 
In $\mathrm{ON}_2$ ($\phi_2=90^\circ$), both modes grow simultaneously and saturate 
at comparable values ($S_y/S_x \sim 1$), confirming square-lattice order. 
The order parameters are quantitatively identical in both protocols, demonstrating 
a fully rewriteable three-state memory.}
\label{fig:SM_protocols}
\end{figure*}

A write pulse with $\Gamma_\mathrm{wr}$ 
drives the photon density transiently far above the ON fixed point, 
erasing the prior supersolid order and relocking to the symmetry 
dictated by $\phi_2$ at the time of the pulse. Consequently, the 
sequence
\begin{equation*}
  \mathrm{OFF}
  \xrightarrow{\mathrm{wr}(0^\circ)} \mathrm{ON}_1
  \xrightarrow{\mathrm{wr}(90^\circ)} \mathrm{ON}_2
  \xrightarrow{\mathrm{wr}(0^\circ)} \mathrm{ON}_1
\end{equation*}
requires only write pulses, with the system remaining on the upper 
branch throughout. Figure~\ref{fig:SM_protocols} (right column) 
demonstrates this protocol. The order parameters $S_x(t)$ and $S_y(t)$ 
are quantitatively identical to those of the write--erase cycles, 
confirming that the write pulse completely resets the supersolid 
symmetry irrespective of the prior state. The three-state device thus 
constitutes a fully rewriteable and bistable memory in which any 
element of $\{\mathrm{OFF},\,\mathrm{ON}_1,\,\mathrm{ON}_2\}$ is 
reachable from any other via at most a single write pulse.

\subsection{Generalisation to $N_L$ layers}

Adding $N_L$ layers at drift angles $\phi_j$ places roton minima at 
the wavevectors $\{\kst\hat{\mathbf{e}}_j\}$ (and their opposites). 
The effective kernel
\begin{equation}
  g_\mathrm{eff}(\mathbf{k})
  = \sum_{j=0}^{N_L-1} g(\mathbf{k};\phi_j) - (N_L-1)\,g_0
  \label{eq:SM_geff_NL}
\end{equation}
satisfies $g_\mathrm{eff}(0)=g_0$ for any configuration, 
so the bistable window and OFF-state thresholds are independent of 
$N_L$. Each layer introduces one additional pair of roton minima of 
identical depth $g(\kst)\approx-0.294$, adding one distinct ON state 
accessible with the same write-pulse conditions (see Sec.~\ref{sec_WPDT}). For equal 
angular spacing ($\phi_j=j\pi/N_L$) all $2N_L$ minima are degenerate, 
and the final ON state is selected by mode competition---a regime 
relevant to non-Abelian switching sequences and controlled pattern 
formation in coupled-cavity arrays.

\bibliographystyle{apsrev4-1}
\bibliography{references_SM}